\documentclass[12pt]{article}

\usepackage{graphicx}
\usepackage{amsmath}
\usepackage{scicite}
\usepackage{times}
\usepackage{rotating}
\usepackage{hyperref}

\newenvironment{sciabstract}{%
\begin{quote} \bf}
{\end{quote}}

\newcommand{\sun}{\odot}
\newcommand{\eg}{\textit{e.g.,}}
\newcommand{\actaa}{Acta Astron.}
\newcommand{\mnras}{Mon. Not. R. Astron. Soc.}
\newcommand{\aj}{Astron. J.}
\newcommand{\aap}{Astron. Astrophys.}
\newcommand{\apj}{Astrophys. J.}
\newcommand{\apjl}{Astrophys. J.}
\newcommand{\apjs}{Astrophys. J. Suppl. Ser.}

\newcommand{\pasp}{Publ. Astron. Soc. Pac.}

\newcommand{\pasj}{Publ. Astron. Soc. Jpn.}
\newcommand{\nat}{Nature}

\newcommand{\araa}{Annual Review of Astronomy and Astrophysics}

\title{A three-dimensional map of the Milky Way using classical Cepheid variable stars} 

\author
{Dorota M. Skowron,$^{1\ast}$ Jan Skowron,$^{1}$ Przemek Mr\'oz,$^1$ Andrzej Udalski,$^{1\ast}$\\
 Pawe\l{} Pietrukowicz,$^1$ Igor Soszy\'nski,$^1$ Micha\l{}~K. Szyma\'nski,$^1$\\
 Rados\l{}aw~Poleski,$^{1,2}$ Szymon Koz\l{}owski,$^1$ Krzysztof Ulaczyk,$^{1,3}$\\
 Krzysztof Rybicki,$^1$ Patryk Iwanek$^1$\\
\\
\normalsize{$^{1}$Astronomical Observatory, University of Warsaw,}\\
\normalsize{Aleje Ujazdowskie 4, 00-478 Warsaw, Poland,}\\
\normalsize{$^{2}$Department of Astronomy, Ohio State University,}\\
\normalsize{140 W. 18th Ave., Columbus, OH 43210, USA}\\
\normalsize{$^{3}$Department of Physics, University of Warwick, Coventry CV4 7AL, UK}\\
\\
\normalsize{$^\ast$E-mail: dszczyg@astrouw.edu.pl, udalski@astrouw.edu.pl}
}

\date{\large {July 18, 2019; submitted: May 29, 2018}}

\begin{document} 

\baselineskip24pt
\maketitle 

\begin{sciabstract} The Milky Way is a barred spiral galaxy, with
physical properties inferred from various tracers informed by the
extrapolation of structures seen in other galaxies. However, the
distances of these tracers are measured indirectly and are
model-dependent. We constructed a map of the Milky Way in
three-dimen\-sions based on the positions and distances of thousands of
classical Cepheid variable stars. This map shows the structure of our
Galaxy's young stellar population, and allows us to constrain the warped
shape of the Milky Way's disk. A simple model of star formation in the
spiral arms reproduces the observed distribution of Cepheids. 
\end{sciabstract}

Cepheid variable stars pulsate, causing their brightness to vary with
periods of 1 to 100 days. Classical Cepheids are young ($<\!400$ million
years old) supergiant stars, whereas other types of Cepheids (type II
and anomalous) arise in older stellar populations. The intrinsic
luminosities of classical Cepheids span the range of 100 to 10,000~solar
luminosities ($L_{\sun}$). This is bright enough to be detected at
extragalactic distances and, within our Galaxy, through obscuring
interstellar clouds of gas and dust in the foreground. Cepheids follow a
pulsation period-luminosity (P-L) relation\cite{leavitt1912}, allowing
the absolute magnitude of a Cepheid to be inferred from its period. The
distance can then be determined by comparing the absolute and the
apparent magnitude, provided the foreground interstellar extinction is
known. At optical wavelengths, interstellar extinction is often large
and highly variable, but observations in infrared (IR) bands reduce
these uncertainties. Detection of Cepheids in the optical bands is
straightforward owing to their characteristic saw-tooth shaped light
curves and large amplitudes. In the IR, the light curve shape becomes
more sinusoidal and difficult to distinguish from other types of
variable stars. Cepheids can appear indistinguishable from other
variable stars in the optical bands if the number of measurements is
limited (below 80 to 100 epochs) and the survey time span is short.

\begin{figure} \centering
\vskip -30pt
\includegraphics[width=0.8\textwidth]{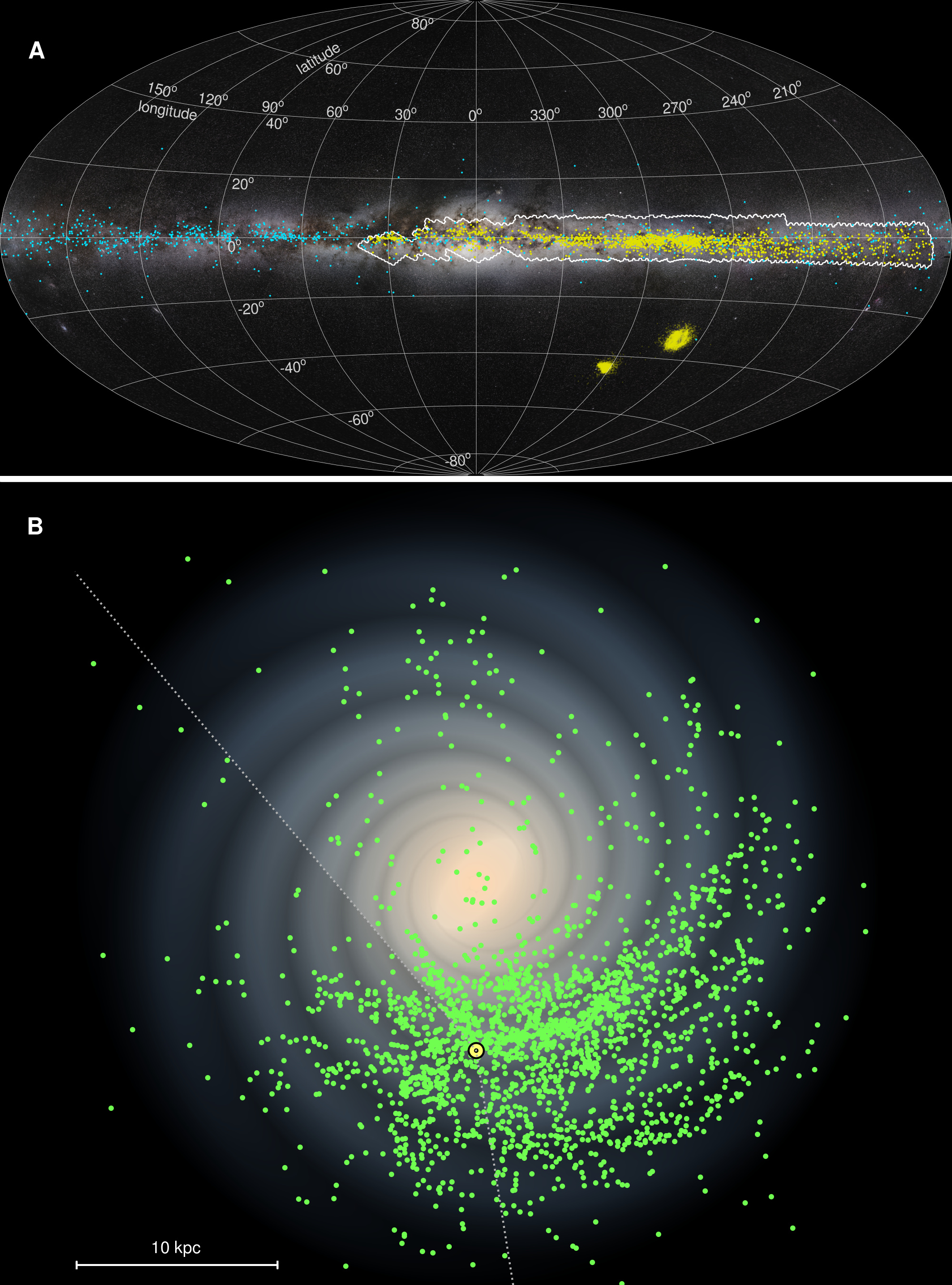}
\caption{\textbf{Distribution of Galactic classical Cepheids.} 
({\bf A}) On-sky view of the Milky Way in Galactic coordinates ($l$,
$b$), with our sample of classical Cepheids in the Milky Way and in the
Magellanic Clouds. Cepheids from the OGLE Collection of Variable Stars
are shown with yellow dots, other sources with cyan dots\cite{suppl}.
The white contour marks the OGLE survey area in the Galactic plane
($190^{\circ} < l < 360^{\circ}, 0^{\circ} < l < 40^{\circ}$;
$-6^{\circ} < b < +6^{\circ}$). The background image is a
Milky Way panorama (by Serge Brunier).
({\bf B}) Face-on view of our Galaxy with all 2431 Cepheids in our
sample marked with green dots. The background image represents a
four-arm spiral galaxy model consistent with neutral hydrogen
measurements in our Galaxy (with the spiral structure modeled as the
logarithmic spirals\cite{vallee2017}). The Sun is marked with a yellow
dot; the dashed lines show the angular extent of the OGLE fields
($190^{\circ} < l < 360^{\circ}, 0^{\circ} < l < 40^{\circ}$).}

\label{fig:distribution}
\end{figure}

We analyzed data for 2431 Galactic Cepheids, most of which were
discovered by the fourth phase of the Optical Gravitational Lensing
Experiment (OGLE-IV) project\cite{udalski2015}. This is a long-term
survey of the Galactic disk and center focused on the discovery and
classification of variable stars. The OGLE Collection of Galactic
Cepheids\cite{udalski2018} more than doubled the number of known
Galactic classical Cepheids. The magnitude range of the Galactic plane 
portion of OGLE-IV, 11 to 18~mag in the {\it I}-band, enables
identification of Cepheids as distant as the expected boundary of the
Galactic disk ($\approx 20\,000$~parsecs (pc) from the Galactic center).
OGLE has covered almost all the Galactic disk visible from its site at
Las Campanas Observatory, Chile (Fig~\ref{fig:distribution}).

\newpage Cepheids closer than 4~kpc are too bright, so they saturate in
the OGLE survey images. Therefore, we complemented the OGLE sample with
brighter objects from the list of Galactic
Cepheids\cite{pietrukowicz2013}, themselves mostly from the General
Catalogue of Variable Stars (GCVS)\cite{samus2017} and the All Sky
Automated Survey (ASAS)\cite{pojmanski2002} survey. We also supplemented
the list with Cepheids discovered by the All-Sky Automated Survey for
Supernovae (ASAS-SN) survey\cite{asassn} and we identified Cepheids in the
Asteroid Terrestrial-impact Last Alert System (ATLAS) survey
catalog\cite{atlas}. We also investigated Cepheid candidates from the
Gaia Data Release 2 (Gaia DR2) catalog\cite{holl2018}, confirming 211 of
them as classical Cepheids. Full details on the sample selection are
provided in \cite{suppl}.

Most of the Cepheids lie close to the Galactic plane, so they are affected by
extinction by dust, which is also concentrated in the plane. To reduce
the effect of dust extinction, we determined the distances to individual
Cepheids based on mid-IR photometry obtained by the Spitzer and
Wide-field Infrared Survey Explorer (WISE) satellites corrected for 
interstellar extinction\cite{suppl}, as well as appropriate P-L
relations\cite{wang2018}. Positions and distances of each Cepheid were
converted to Cartesian coordinates with the Sun at the origin to
facilitate study of their 3-D distribution \cite{suppl}.

Fig.~\ref{fig:distribution} presents our map of the Galaxy in Cepheids,
representing the young stellar population. We show the view projected on
the sky (Fig.~\ref{fig:distribution}\,A) and face-on 
(Fig.~\ref{fig:distribution}\,B), together with a four-arm spiral galaxy
model consistent with neutral hydrogen (H~\textsc{i})
observations\cite{levine2006}. Areas of higher Cepheid density indicate
the non-uniform Galactic structure or regions of enhanced star
formation. The side view (Fig.~\ref{fig:distribution}\,A) shows that the
young stellar disk of the Milky Way between longitudes $240^{\circ} < l
< 330^{\circ}$ lies below the Galactic plane (as traced by Cepheids),
suggests warping of the Galactic disk.

We subdivided the Galaxy into 12 sectors of unequal azimuthal width in
the Galactocentric polar coordinate system with the azimuth
$\phi=0^{\circ}$ pointing to the Sun (Fig.~\ref{fig:warp}\,A). The disk
is not flat -- the warp is evident in directions that are sufficiently
well populated with Cepheids (Fig.~\ref{fig:warp}\,B). The warping of
the disk begins at a distance of $\sim 8$~kpc and becomes steeper at
$\sim 10$~kpc, reaching out to the edge of the Galaxy at $\sim 20$~kpc.
The disk warps toward negative distance from the Galactic plane ($Z$) in
the azimuth range $0^{\circ} < \phi < 135^{\circ}$ and toward positive
$Z$ from $165^{\circ} < \phi < 330^{\circ}$, although the exact boundary
is unclear due to the lower number of known Cepheids on the far side of
the Galactic center. To further illustrate the warping of the disk and
as a guide to the eye, we fitted a simple model surface to the Cepheid
distribution\cite{suppl}, which is shown in Fig.~\ref{fig:warp}\,C to E
from three viewing angles.

The warping of the disk has been observed before, in
H~\textsc{i}\cite{nakanishi2003,levine2006},
stars\cite{drimmel2001,lopez2002,momany2006,reyle2009,amores2017},
dust\cite{marshall2006}, and kinematics of stars in the plane of the
sky\cite{smart1998,poggio2018}. Our map shows the warp in three
dimensions and differs from models derived from those earlier
detections\cite{suppl}.

We modeled the thickness of the young disk by fitting a simple
exponential model\cite{suppl}. We measured the disk scale height within
the solar orbit $H=73.5\pm3.2$~pc and the distance of the Sun from the
Galactic plane $z_0=14.5\pm3.0$~pc (Fig.~\ref{fig:hist}). These values
can be compared to the past determinations (Table~\ref{tab:h_z_fit}). We
removed the average warp shape from the  Galactic disk to analyze its
flaring properties far from the center (Fig.~\ref{fig:flaring}). We found
that the flaring of the disk in Cepheids matches that seen in neutral
hydrogen\cite{kalberla2007} (Fig.~\ref{fig:flaring}\,B).

The distribution of classical Cepheid pulsation periods as a function of
the distance from the Galactic center is presented in
Fig.~\ref{fig:per-dist}. It shows a decrease in the minimal pulsation
periods of Cepheids with the increasing Galactocentric distance
consistent with a radial gradient of metallicity (abundance of elements
heavier than hydrogen and helium) in the Milky
Way\cite{antonello2002,anderson2016,genovali2014}.

The age of classical Cepheids is correlated with their pulsation period,
metallicity and stellar rotation\cite{anderson2016}. Using the period
distribution, we performed an age tomography of the Milky Way Cepheids
(Fig.~\ref{fig:ages}\,A).  After applying the period-age relations for
Cepheids\cite{anderson2016} and including the metallicity
gradient\cite{genovali2014,suppl}, we determined that the majority of
Cepheids in our sample formed between 50 million and 250 million years
ago (Fig.~\ref{fig:ages}\,B). The spatial distribution of ages shows
that the closer to the Galactic center, the younger Cepheids we observe
(Fig.~\ref{fig:ages}\,C to E).  The age distribution of Cepheids does
not directly reflect that of all stars present in the disk. For example,
the absence of short-period Cepheids in the inner disk does not
necessarily mean that older (less massive) stars are absent in that
region, as the higher metallicity there would not have produced Cepheids
in the older population.

The spatial distribution of Cepheids shows several distinct features
(Fig.~\ref{fig:ages}\,A). Because these features are located mainly in
the area monitored by OGLE, where the Cepheid detection efficiency is
high\cite{udalski2018}, they are most likely real, rather than the
result of an observational bias. The most prominent feature is formed by
Cepheids within an age range of 90 to 140 Myr (Fig.~\ref{fig:ages}\,D).
Traces of this arc-shaped overdensity were previously associated with
the Sagittarius-Carina spiral
arm\cite{majaess2009,dambis2015,sanna2017,vallee2017}. An additional
small overdensity in this age range is connected with the Perseus arm.

Fig.~\ref{fig:distribution}\,B also shows additional overdensities of
Cepheids. In Fig.~\ref{fig:ages}\,C, E we present the distribution of
our Cepheids in the age ranges of 20 to 90 Myr and 140 to 260 Myr,
respectively. There are three overdensities in the youngest bin and two
in the oldest one. The youngest overdensities  roughly correspond to the
inner Galaxy spiral arms (Norma-Cygnus, Scutum-Crux-Centaurus, and
Sagittarius-Carina) whereas for the oldest ones, possible associations
with the Perseus and Norma-Cygnus/Outer arms are much less evident.
Cepheids within individual overdensities have similar ages, suggesting a
common origin in past star formation episodes.

We performed a simple simulation in which we select stars from three
age bins. Median age values in these three groups were 64, 113 and 175
Myr. Then we searched for a star formation episode in one or more of the
spiral arms, which, after a given time (after the Galaxy has rotated),
would produce the currently observed distribution of Cepheids. The
results of this simulation are presented in Fig.~\ref{fig:rotation},
where each row focuses on one of the age bins.

Fig.~\ref{fig:rotation}, A, D, and G, shows the current observed
distribution of Cepheids. Fig.~\ref{fig:rotation}, B, E, and H presents
the positions of the spiral arms as they were 64 Myr, 113 Myr, and 175
Myr ago, respectively, following an assumed rotational period of the
spiral pattern equal to 250 Myr\cite{vallee2017}. At those moments and
locations, we injected star formation episodes. Fig.~\ref{fig:rotation},
C, F, and I, shows how those star formation regions would look now
(after 64, 113, and 175 Myr, respectively), taking into account the
typical velocity of disk stars\cite{mroz2018}. We require a low-velocity
dispersion of Cepheids during their birth (8 km/s) to match the coherent
overdensities we observe.

Even with these simple assumptions, we find a good match between the
observed and simulated distributions of Cepheids in the shape of the
overdensities, as well as in their internal dispersion. Cepheids that
were formed in a spiral arm, do not currently follow the exact location
of that arm; this can be explained by the difference in rotation
velocity  between the spiral density waves and the stars. This is the
most pronounced in the case of the oldest group
(Fig.~\ref{fig:rotation}\,G to I), where the overdensities fall between
the Perseus and Norma-Cygnus/Outer arms, in which they were most likely
formed, and in the Sagittarius-Carina -- Perseus arm gap.

We have produced a three-dimensional map of the Milky Way based on a
large number of individual Cepheids with measured distances. This
represents the young stellar population that extends to about 20~kpc,
covering a large portion of our Galaxy, thus illustrating the extent and
shape of the young stellar disk. Our work shows that a simple model can
reproduce the current distribution of the young stellar disk of the
Milky Way with narrow patches of stars of similar age. 

\newpage


\bibliographystyle{Science}

\section*{Acknowledgments}

We would like to thank the Reviewers for constructive comments that
helped to improve the paper. This publication makes use of data products
from the Wide-field Infrared Survey Explorer, which is a joint project
of the University of California, Los Angeles, and the Jet Propulsion
Laboratory/California Institute of Technology, funded by the NASA. This
work is based in part on observations made with the Spitzer Space
Telescope, which is operated by the Jet Propulsion Laboratory,
California Institute of Technology under a contract with NASA.

{\bf Funding:} 
The OGLE project has received funding from the National Science Center
(NCN), Poland through grant MAESTRO 2014/14/A/ST9/00121 (A.U.). Also
supported by NCN grant 2013/11/D/ST9/03445 (D.M.S.), NCN grant MAESTRO
2016/22/A/ST9/00009 (I.S.) and the Foundation for Polish Science --
Program START (P.M.).
{\bf Authors contributions:} D.M.S., J.S. and P.M. analyzed, modeled and
interpreted the Galactic Cepheid data. A.U. initiated, supervised the
project and prepared the sample of Cepheids. P.P. and I.S. made the
final verification of the sample used in this analysis. M.K.S. provided
positional data for Cepheid variables. D.M.S., J.S., P.M. and A.U.
prepared the manuscript. All authors collected the OGLE photometric
observations and reviewed, discussed and commented on the results and on
the manuscript.
{\bf Competing interests:} The authors of the manuscripts have no competing
interests.
{\bf Data and materials availability:} Data for full Cepheid sample is
available in Table~S1 and Data~S1. The OGLE Collection of Galactic
Cepheids is available at the OGLE Archive:
\begin{center}
{\it http://www.astrouw.edu.pl/ogle/ogle4/OCVS/}
\end{center}
The Python code for the simulations shown in Fig.~\ref{fig:rotation} is
available at
\begin{center}
{\it https://github.com/jskowron/galactic\_cepheids}
\end{center}

\vspace*{10mm}
\section*{Supplementary Materials}
Materials and Methods\\
Figs. S1--S5\\
Tables S1--S2\\
Data S1\\
References \textit{(31--69)}\\
\\
29 May 2018; accepted 5 July 2019\\
10.1126/science.aau3181

\newpage

\newpage

\begin{figure}
\centering \includegraphics[width=1.0\textwidth]{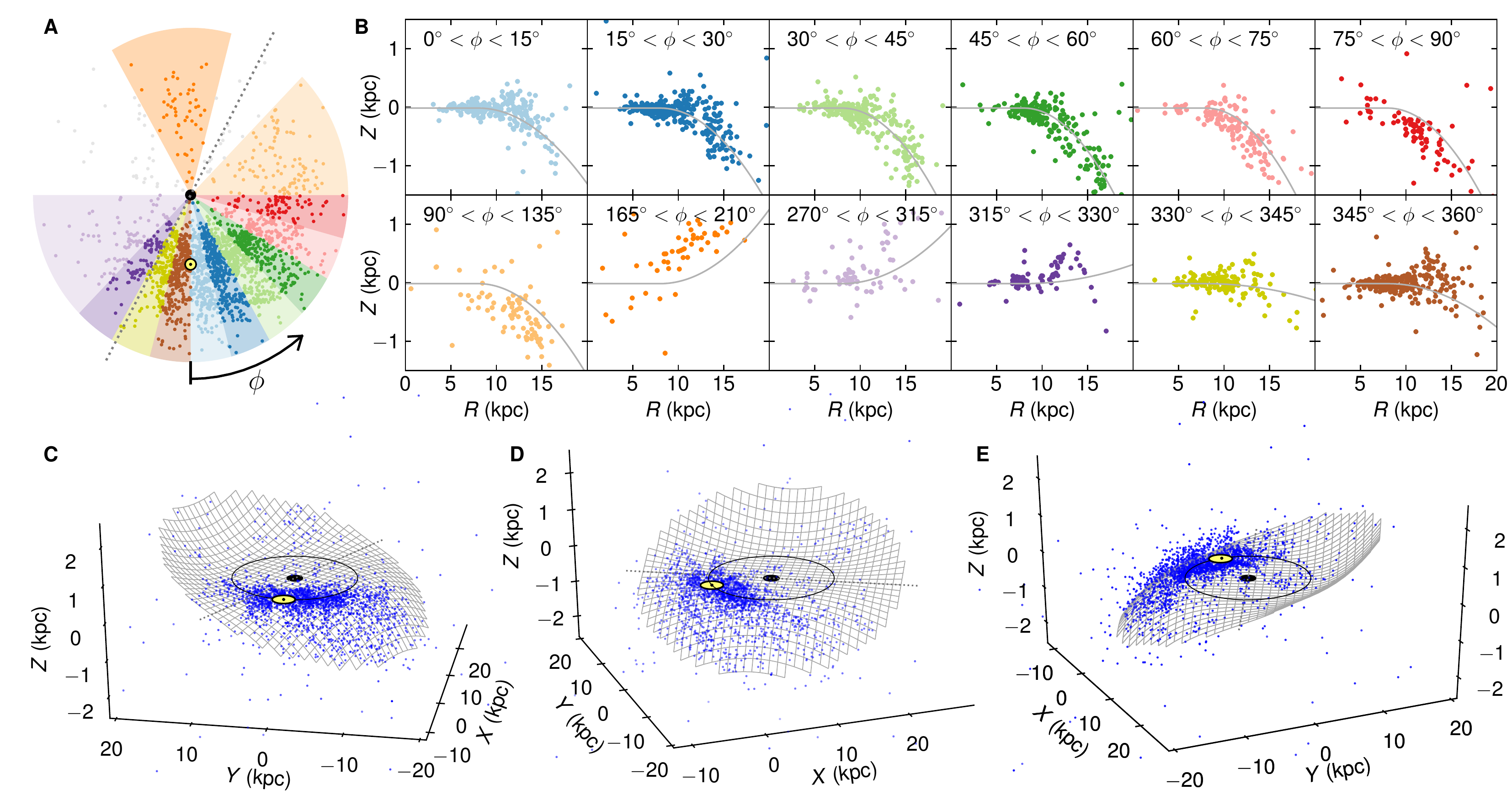}

\caption{\textbf{Warping of the Milky Way's disk.} 
({\bf A}) Top view of the Galactic disk divided into 12 sectors in the
Galactocentric polar coordinate system. The azimuth $\phi=0^\circ$ is
pointing to the Sun. The Sun is marked with a yellow dot, the Galactic
center with a black dot. Each sector is shown with a different color and
is analyzed in panel (B). The dotted line on the disk separates parts of
the Galaxy warped toward negative and positive $Z$ (i.e., the line of
nodes). 
({\bf B}) Distribution of Cepheids as a function of the Galactocentric
distance $R$ versus distance from the Galactic plane $Z$ in each sector.
The range of azimuth for each sector is labeled. Colors correspond to
colors of the sectors in (A). Gray lines are positions of the model
surface ((C) to (E)) within a given sector. 
({\bf C} to {\bf E}) Model of the Milky Way warp from three viewing
angles. Cepheids are marked with blue dots, the gray grid is a model
surface fit to the Cepheid distribution\cite{suppl}. Cartesian Galactic
coordinate system ($X, Y, Z$) is defined in \cite{suppl}. Vertical
distances have been exaggerated by the choice of axes.}
\label{fig:warp}
\end{figure}

\newpage

\begin{figure}
\centering \includegraphics[width=1.0\textwidth]{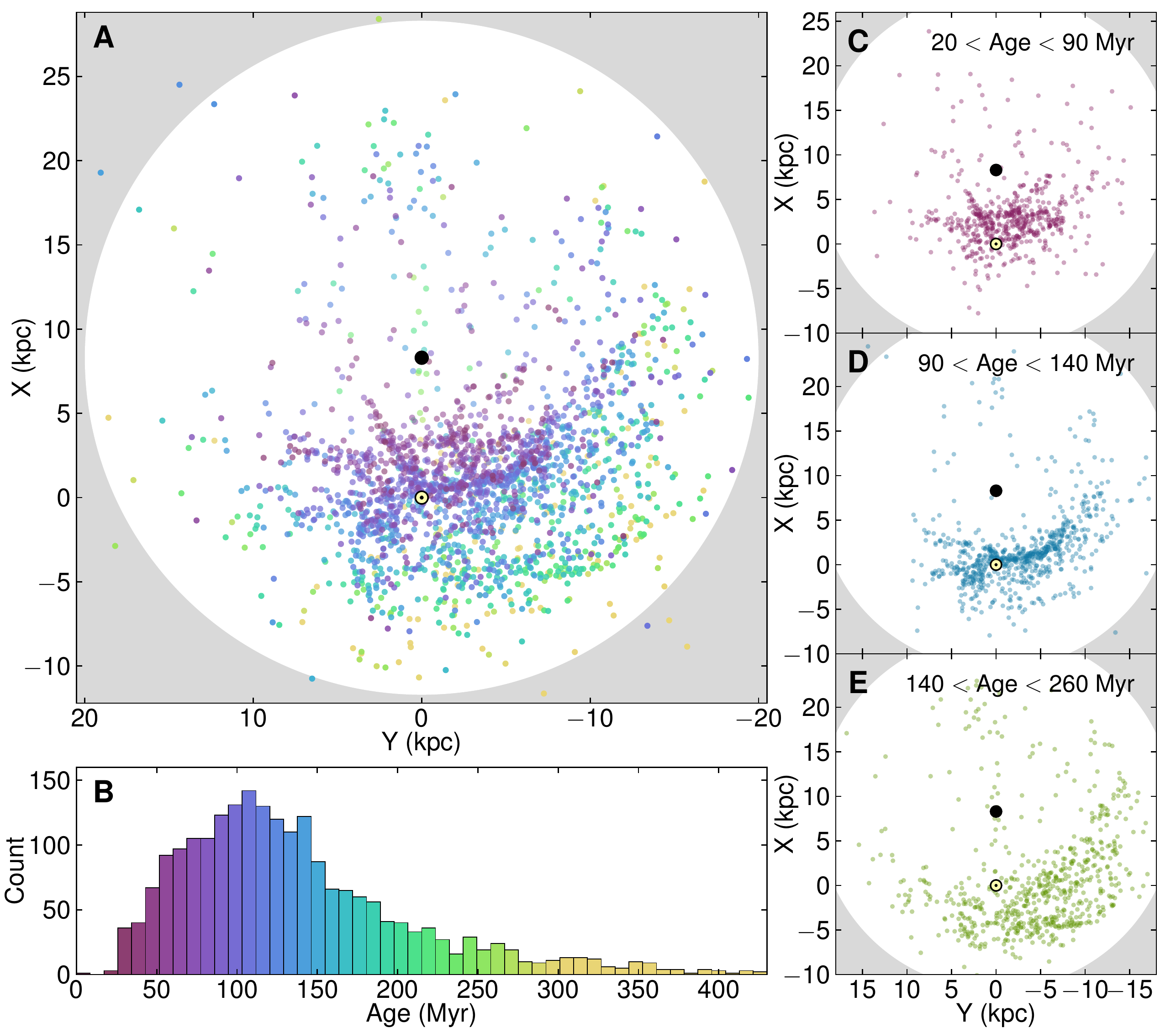}
\caption{\textbf{Ages of Galactic classical Cepheids.}
({\bf A}) Face-on view, showing the Cepheid distribution in the Galaxy,
with colors corresponding to Cepheid ages as indicated in (B). The Sun
is marked with a yellow dot, the Galactic center with a black dot.
({\bf B}) Age histogram of Galactic classical Cepheids in our sample.
({\bf C} to {\bf E}) Age tomography of the Milky Way Cepheids in three
selected age bins, as indicated. Each age bin reveals Cepheid
overdensities.}
\label{fig:ages}
\end{figure}

\newpage

\begin{figure}
\centering \includegraphics[width=0.8\textwidth]{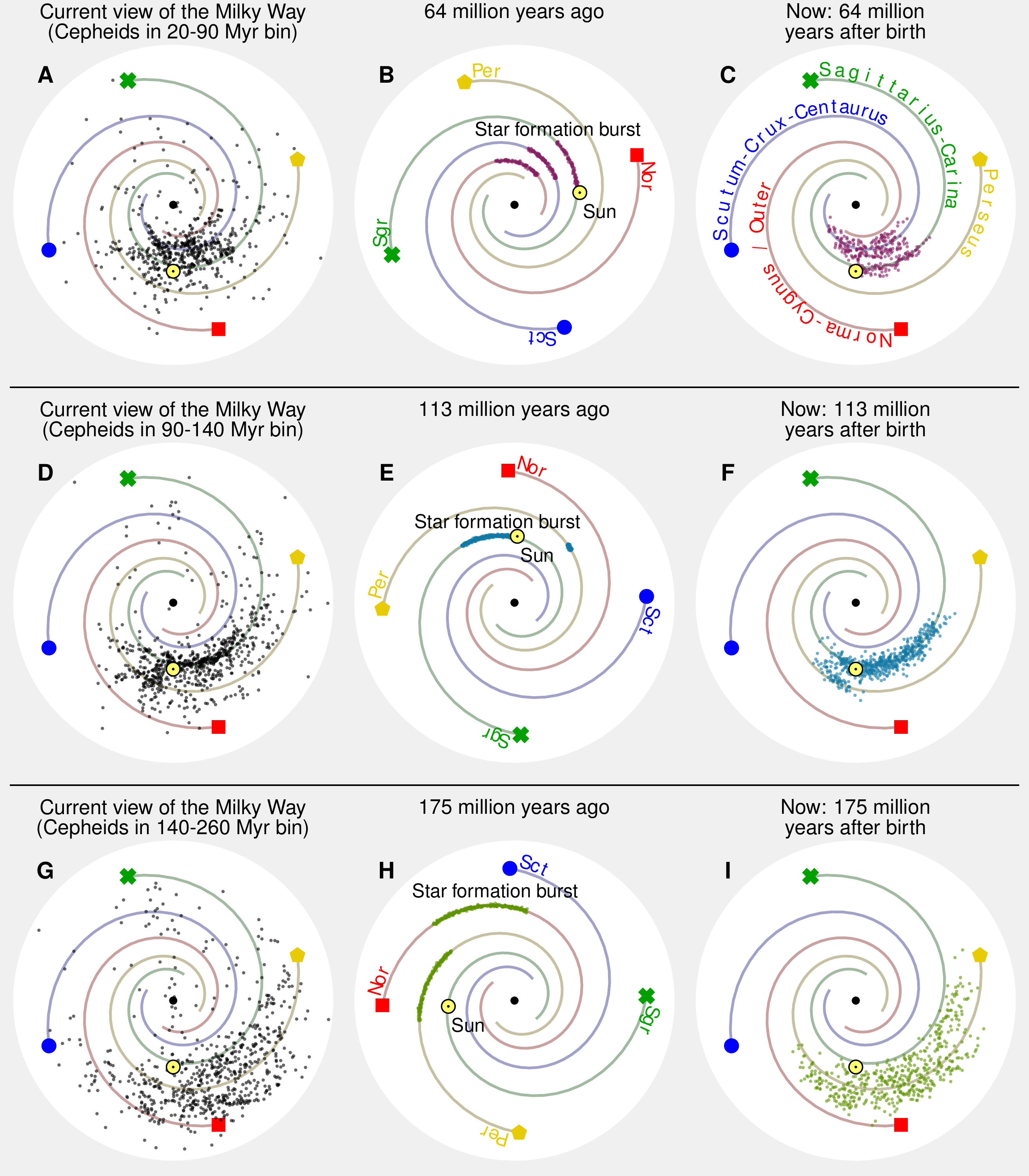}
\caption{\textbf{Possible origin of the Cepheid structures.}
({\bf A}) Face-on view of our Galaxy, where Cepheids that belong to the
age bin 20 to 90~Myr (median age $64$~Myr) are shown with black dots.
The Sun is marked with a yellow dot, the Galactic center with a black
dot. Locations of the spiral arms: yellow pentagon, Perseus arm; green
cross, Sagittarius-Carina arm; blue dot, Scutum-Crux-Centaurus arm; red
square, Norma-Cygnus/Outer arm.
({\bf B}) Location of the Galaxy's spiral arms $64$~Myr ago, with
simulated star formation regions along the Norma-Cygnus/Outer,
Scutum-Crux-Centaurus and Sagittarius-Carina arms marked in violet.
({\bf C}) Current location of stars from the simulated star formation
region (violet).
({\bf D} to {\bf F}) Same as (A) to (C), but for the age bin 90 to
140~Myr with a median age of $113$~Myr.
({\bf G} {\bf I}) Same as (A) to (C), but for the age bin 140 to 260~Myr
with a median age of $175$~Myr.}
\label{fig:rotation}
\end{figure}

\clearpage


\newpage
\renewcommand\thefigure{S\arabic{figure}}
\setcounter{figure}{0}
\renewcommand\thetable{S\arabic{table}}
\renewcommand{\theequation}{S\arabic{equation}}
\pagenumbering{arabic}
\setcounter{page}{1}

\begin{figure} \centering
\includegraphics[width=0.4\textwidth]{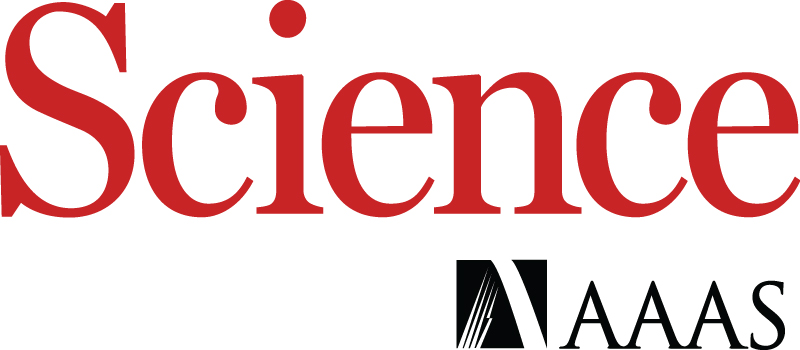}
\end{figure}

\vspace*{10mm}
\section*{\centerline{\Large Supplementary Materials for}}

\bigskip
\title{\centerline{\Large A three-dimensional map of the Milky Way}}
\title{\centerline{\Large using classical Cepheid variable stars}} 
\bigskip

\begin{center}
Dorota M. Skowron, Jan Skowron, Przemek Mr\'oz, Andrzej Udalski,
Pawe\l{} Pietrukowicz, Igor Soszy\'nski, Micha\l{}~K. Szyma\'nski,
Rados\l{}aw~Poleski, Szymon Koz\l{}owski, Krzysztof Ulaczyk,
Krzysztof Rybicki, Patryk Iwanek\\

\vskip 1cm
\normalsize{Correspondence to: dszczyg@astrouw.edu.pl, udalski@astrouw.edu.pl}

\end{center}

\noindent
{\bf This PDF file includes:}

\noindent
\hglue0.5cm Materials and Methods\\
\hglue0.5cm Figs. S1 to S5\\
\hglue0.5cm Tables S1 to S2\\

\noindent
{\bf Other Supplementary Materials for this manuscript include the following:}

\noindent
\hglue0.5cm Data S1\\

\newpage

\section*{~}

\subsection*{Materials and methods}

\bigskip
\bigskip
\subsubsection*{The Cepheid sample}
 
The sample of classical Cepheids used in this work consists of 2431
objects. The majority (1514 stars) come from the OGLE Collection of
Galactic Cepheids\cite{udalski2018} (1408 objects from the OGLE-IV
Galactic disk and outer Galactic bulge fields\cite{udalski2018},
68~objects from the OGLE-IV collection of Galactic bulge
Cepheids\cite{soszynski2017}, and 38 objects from the OGLE-III catalog
of variable stars in the Galactic bulge\cite{soszynski2011} and
disk\cite{pietrukowicz2013}). Some of these objects have counterparts in
other databases but the majority do not. We also supplement this list
with brighter all-sky Cepheids: 592 from the General Catalogue of
Variable Stars (GCVS)\cite{samus2017}, 61 from the ASAS
catalogs\cite{pojmanski2002}, and 15 from other sources (Data~S1).

The catalog of new variable stars from the ASAS-SN survey\cite{asassn}
contains 315 objects classified as classical (249) or type II (66)
Cepheids, together with their V-band light curves. Because the
classification procedure of ASAS-SN is mainly automatic, we visually
inspected all Cepheid light curves. As a result, we confirmed 147
classical and 43 type II ASAS-SN Cepheids, with a misclassification rate
between the original Cepheid types of about 30\%. The remaining variable
stars are mainly spotted variables, eclipsing binaries, anomalous
Cepheids, or RR Lyr type stars. Among 147 classical Cepheids, 50 were
already in our sample, so we supplement it with 97 ASAS-SN objects.

We also used the ATLAS project catalog of variable stars\cite{atlas} to
identify classical Cepheids. Following the survey documentation, we
selected all stars with the 'PULSE' classification that have a
long-period fit or a short-period fit leading to a master period longer
than 1 day. We also selected stars that were not classified as 'PULSE'
but reveal clear variability, and their master period is longer than 1
day. See \cite{atlas} for further details. The search resulted in almost
3000 Cepheid candidates with available light curves. Our analysis of the
ATLAS time series was similar to the procedures applied to the OGLE
photometry. We determined periods, amplitudes, and Fourier coefficients
for all available light curves in both passbands (cyan and orange) used
by the ATLAS survey. Then, we selected and classified pulsating stars
based on the visual inspection of their folded light curves, taking into
account their position in the Period--Fourier coefficients
diagrams\cite{soszynski2008,soszynski2017} and ratios of the light
amplitudes in the two ATLAS filters. The classification procedure
returned 438 classical Cepheids, 152 of which were not already in our
sample. We also identified 142 type II Cepheids and 61 anomalous
Cepheids in the ATLAS Cepheid candidate sample. The majority of the
remaining 2130 ATLAS objects turned out to be eclipsing binaries and
spotted variables, and even some RR Lyrae type stars that were assigned
a 1-day alias period.

The OGLE Cepheid Collection confirms 211 classical Cepheid candidates
published in the Gaia DR2 Cepheid sample\cite{holl2018}. Four of them 
were not in the original OGLE catalog\cite{udalski2018} but are
confirmed with OGLE-IV data. Galactic Gaia DR2 Cepheid candidates
located outside of the OGLE sky footprint were not included in our
sample due to contamination and low completeness\cite{udalski2018}.

\begin{figure}[htb]
\centering \includegraphics[width=0.85\textwidth]{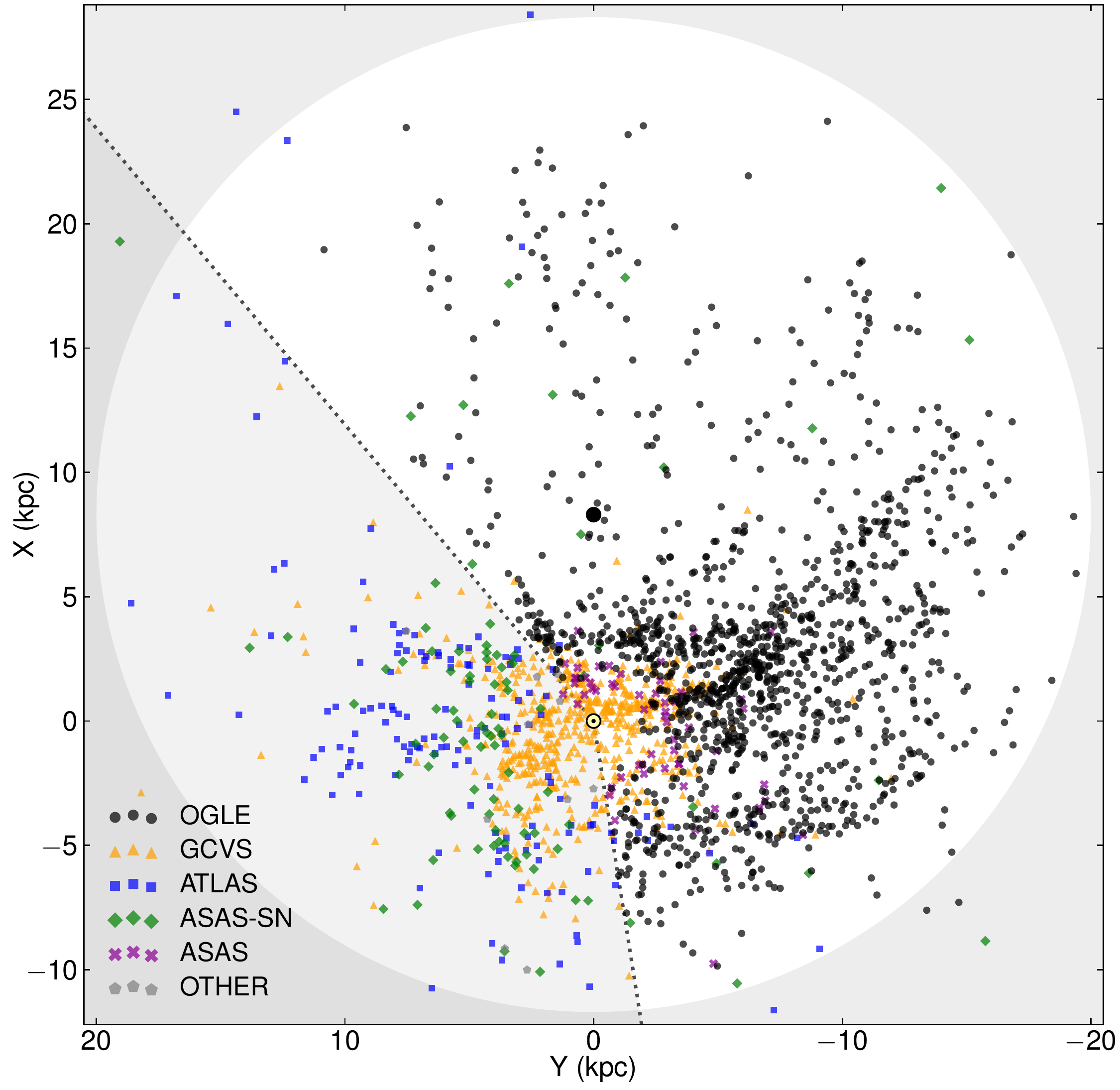}
\caption{\textbf{Distribution of Galactic classical Cepheids from
different databases.} Same as Fig.~\ref{fig:distribution}\ B, but
indicating the source of each Cepheid on a plain background. See text
for details of the sources.}
\label{fig:distribution_source}
\end{figure}

\begin{figure}[htb]
\centering \includegraphics[width=0.9\textwidth]{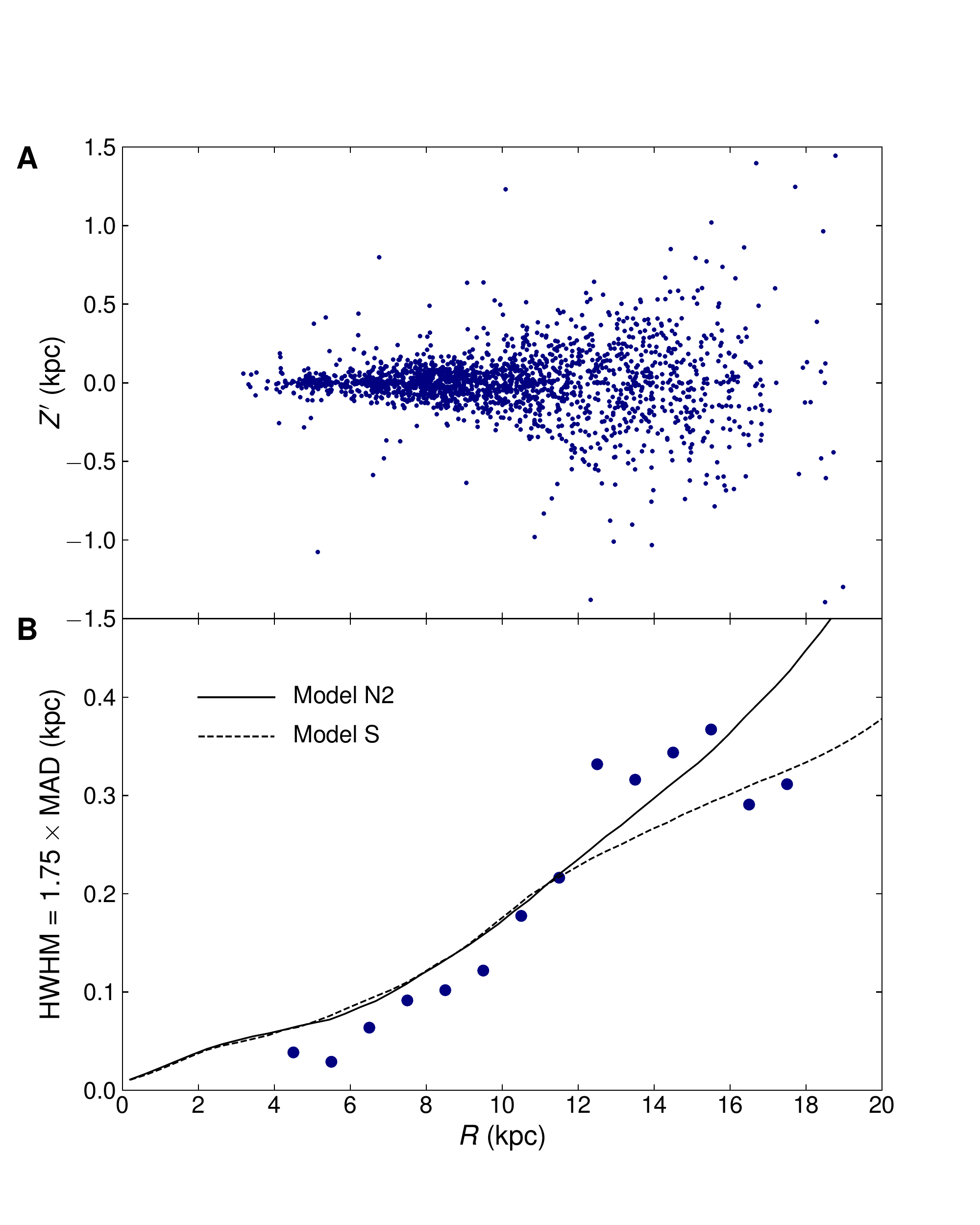}
\caption{\textbf{Flaring of the Galactic disk in classical Cepheids.}
{\bf (A)}: Cross-section of the Galactic disk after subtracting the
median warping ($Z'=Z - $~median warping).
{\bf (B)}: Thickness of the flaring disk as a function of the
Galactocentric radius (MAD is the median absolute deviation and HWHM the
half-width at half-maximum). Dots represent the investigated Cepheid
sample. Solid and dashed lines show models of flaring of the northern
(N2) and southern (S) part of the disk, based on radio observations of
atomic hydrogen\cite{kalberla2007}.}
\label{fig:flaring}
\end{figure}

The full list of 2431 classical Cepheids used in this work is presented
in Table~\ref{tab:main}, including a reference to a discovery paper,
whenever available. Otherwise we provide the source (\eg~GCVS) or a
database, in which the Cepheid was identified (\eg~ATLAS). For Cepheids
from the OGLE Collection that have counterparts in other databases we
provide an additional database identifier.
Fig.~\ref{fig:distribution_source} presents the distribution of our
Cepheid sample in the Milky Way, coded by the source of data.

\subsubsection*{MID-IR data}

The mid-IR observations of Cepheids were taken by the Wide-field
Infrared Survey Explorer (WISE) and the Spitzer Space Telescope.

We utilized the WISE data from the AllWISE Multiepoch Photometry
Database, which provides all-sky time-series photometry from both the
WISE cryogenic and ``Near Earth Object WISE'' (NEOWISE) post-cryogenic
phases of the survey\cite{wise,neowise}. The observations were made in
four bands: W1 (3.4~micron), W2 (4.6 micron), W3 (11.6 micron), and W4
(22.1 micron). We found counterparts for 2068 out of 2431 Cepheids
within $1''$ search radius. The number of points per light curve in the
W1 band varied from 0 to over 301 with a median value of 40 points. We
calculated weighted mean magnitudes for 2041, 2055, 1488 and 524
Cepheids in W1, W2, W3 and W4 bands, respectively.

While calculating the weighted mean magnitude, the weight of each data
point is taken as an inverse square of the uncertainty reported by the
survey ($\sigma_i^{-2}$). We assign the uncertainty of the weighted mean
by taking three factors into account:  the weighted standard deviation
of the light curve, the standard error of the weighted mean calculated
from the weights ${(\sum_i \sigma_i^{-2})}^{-1/2}$, and the expected
root mean square (rms) scatter of the Cepheid light curve, caused by its
pulsations. We found that the uncertainties reported in the catalog are
not a reliable indication of the true observational uncertainty in this
dataset.  The WISE light curve typically covers only a small portion of
the pulsational period. Hence, typically, its standard deviation better
corresponds to the observational errors than to the pulsational scatter.
To be conservative, we decided to use either the weighted standard
deviation of the measurements or the standard error of the weighted
mean, whichever is greater, and combine it in quadrature with the
expected uncertainty originating from  pulsations.  Using this approach,
at the cost of possible minor overestimation of the uncertainty (up to a
factor of $\sqrt{2}$), we avoid overconfidence in the mean brightness
measurement for many stars.  

Since the mid-IR amplitudes for the majority of Cepheids are not known,
due to the low number of observations, we use their optical light curves
to estimate the expected mid-IR amplitude. Most Cepheids in our sample
have well covered {\it I}-band light curves, so their amplitude and thus
the expected rms scatter can be estimated. In the case when only the
{\it V}-band light curve is available, we scale the {\it V}-band
amplitude by a factor of 0.6 to estimate the {\it I}-band amplitude and
rms scatter\cite{klagyivik2009}.

We then choose a subset of Cepheids with well sampled light curves for
every WISE band. From those, we pick the ones with low photometric noise
and estimate their mid-IR rms scatter. We found that  the rms scatter in
the {\it I}-band is proportional to the rms scatter in all mid-IR bands
with the proportionality constant close to 2. Therefore, we take the
measured rms {\it I}-band scatter, divide it by 2, and use it as a
measure of the mid-IR rms scatter.

The measurements from WISE are often grouped in time, as such, all
measurements collected in close temporal proximity can be treated as a
single-epoch. The estimation of a mean brightness having only one or two
measurements across all pulsation phases is uncertain. When we assume
that the pulsational light curve is sinusoidal in shape, the uncertainty
of the mean as estimated from a single-epoch measurement can be taken as
0.7 of the rms scatter, and as estimated from two epochs -- 0.5 of it.

Table~\ref{tab:main} contains our assessment of the mean magnitudes  in
all four WISE bands together with the uncertainties we assigned to these
measurements.

We also use the data from the ``Galactic Legacy Infrared Midplane Survey
Extraordinaire''\linebreak[4]  (GLIMPSE) Legacy Program of Spitzer and
its extensions: GLIMPSE II, GLIMPSE 3D,\linebreak[4] GLIMPSE 360, Deep
GLIMPSE and Vela-Carina, as well as from the "Spitzer Mapping of the
Outer Galaxy" (SMOG) and "A Spitzer Legacy Survey of the Cygnus-X
Complex" (Cygnus-X) Legacy Programs\cite{glimpseI,glimpse}, which we
will collectively call the GLIMPSE data. The Spitzer observations of the
Galactic plane were made in four Infrared Array Camera (IRAC) bands: I1
(3.6~micron), I2 (4.5~micron), I3 (5.8~micron) and I4 (8.0~micron). For
each Cepheid from our final list we extracted all available GLIMPSE
observations within $1''$ search radius. We found matches for 1247 out
of 2431 objects. The majority of Cepheids (980) had only one match in
the GLIMPSE catalogs, while 267 were found in more than one GLIMPSE
programs: 234 stars had 2 matches, 31 stars had 3 matches and 2 stars
had 4 matches. We calculated mean magnitudes for 1121, 1154, 772 and 731
Cepheids in I1, I2, I3 and I4 bands, respectively.  

The number of data points in each light curve is very limited, and the 
majority of objects have one or two measurements. For objects with only
one measurement, we take its value as an approximation of the mean
magnitude and attach the uncertainty based on two factors: the reported
uncertainty of the data point, and the expected rms scatter based on the
pulsation amplitude multiplied by 0.7. For objects with two or more data
points, we use the standard error of the weighted mean and the rms of
pulsation multiplied by 0.5. The results are presented in
Table~\ref{tab:main}.

\bigskip
\bigskip
\subsubsection*{Distances and extinction}

The distance $d$ in a given band $\lambda$ can be calculated as: 
\begin{equation} 
d_{\lambda} = 10^{0.2(m_{\lambda}-M_{\lambda}-A_{\lambda})+1}~{\rm pc} 
\end{equation}
where $m_{\lambda}$ and $M_{\lambda}$ are, respectively, the observed
and absolute magnitudes of the star and $A_{\lambda}$ is the extinction
value. The absolute magnitudes can be calculated from the P-L relations.
Here we use the mid-IR P-L relations derived for the Spitzer and WISE
passbands\cite{wang2018} based on a sample of 288 Galactic classical
Cepheids. The extinction values in the Galaxy are high in the optical
bands and vary strongly between lines-of-sight, but become much lower in
the mid-IR. Nevertheless, in some directions, especially close to the
Galactic plane or in the Galactic bulge, the mid-IR extinction can be
large. Therefore, we assess its value for each individual Cepheid.

We used the 3-D map of interstellar extinction ``mwdust''\cite{bovy2016}
that provides extinction in the requested band (in our case $K_s$) for
the Galactic coordinates ($l$, $b$) and the distance $d$ from the Sun.
The map contains two merged 3-D maps. For the highly extincted part of
the Galaxy ($-100^{\circ} < l < 100^{\circ}$) it uses the map of
ref.\cite{marshall2006} based on ``Two Micron All-Sky Survey'' (2MASS),
and for the remaining longitudes the Bayestar15 map\cite{green2015}
based on ``Panoramic Survey Telescope And Rapid Response System''
(Pan-STARRS) and 2MASS. The former map samples the extinction up to
12--15~kpc from the Sun. The vast majority of Cepheids are located
within that range, but for single objects the extinction may be slightly
underestimated because of neglecting residual reddening in the outskirts
of the Galaxy. The extinction map of ref.\cite{green2015} has a shorter
range (5--6~kpc), but it is used in the directions of lower reddening,
where Cepheids are located much closer. Again, the extinction may be
slightly underestimated for some Cepheids located farther than 6~kpc in
these directions.

To make sure that our IR extinction values from ``mwdust'' maps are
reliable, especially in the lines-of-sight with higher extinction, we
conducted an additional test. We tried to derive $A_{K_s}$ for each
individual Cepheid using its multiband IR photometry. The mid-IR
extinction curve is fairly flat, so the mid-IR photometry alone is 
insufficient for this purpose, and at least a $K_s$-band measurement is
necessary to have an estimate of $A_{K_s}$. We used the ``VISTA
Variables in the Via Lactea'' (VVV) survey
data\cite{saito2010} and extracted $K_s$-band photometry for a sample of
our Cepheids. For the objects having more than 7 epochs, we calculated
their mean intensities. Then assuming the mid-IR extinction
curve\cite{xue2016} (i.e., extinction ratios $R_i$ = $A_i$ / $A_{K_s}$)
and the mid-IR photometry we derived the most likely extinction
$A_{K_s}$ and extinction-corrected distance modulus $\mu_0$, by solving
a set of linear equations $(m-M)_i = \mu_0 + R_i \times A_{K_s}$, for
273 objects. The accuracy of this measurement strongly depends on the
$K_s$ value obtained from the VVV aperture photometry. 

\newpage
Generally, these $A_{K_s}$ values and the ones from ``mwdust'' are
consistent, which is reassuring. On average, the ``mwdust'' extinction
value is only slightly lower (by 0.05~mag) albeit the dispersion of
$A_{K_s}$ measurements (0.5~mag) is large. Because of the strong
dependence of the accuracy of the extinction determination on the
Cepheid $K_s$ photometry, we conclude that the ``mwdust'' extinction
value is more reliable and homogeneous and thus, we use it in the
distance determination.

The distances to each object can be derived separately for every
available Spitzer and WISE band and then averaged. We take the
extinction-corrected distance moduli ($\mu_{0,\lambda} =
m_{\lambda}-M_{\lambda}-A_{\lambda}$) and calculate their weighted
average with 3-$\sigma$ outlier rejection, and from this calculate the
final distance. The mean-brightness uncertainty is added to the relevant
P-L relation scatter\cite{wang2018} in quadrature and the weights are
set as inverse squares of these values. The standard deviation of the 
weighted mean of the distance modulus ($\sigma_\mu$) propagates to the 
final distance uncertainty as $\sigma_d = d \times ln(10) \times 0.2 
\sigma_{\mu}$. 

When averaging the distances, we found that some portion of rejected
measurements comes from the W3 and W4 bands that yield systematically
lower distances. Stars for which this happens often have  complete set
of good measurements in shorter wavelengths (W1 or W2)  from all WISE
visits, but only a handful of measurements in longer wavelengths (W3 or
W4). We take this as an indication of an unreliable measurement in a
given band and discard it whenever the number of useful epochs is
smaller than 20\% of all visits. This procedure lowers the number of
outliers we have to reject and helps to better estimate distances for
both faint, distant  stars (with unreliable W3 and W4 measurements), and
also to bright, close-by stars, which are saturated in shorter mid-IR
bands, and their distances are usually determined solely from the W3 and
W4 bands.

The final distance and the estimated extinction are correlated. We use
simple iterations to converge on the final values. First, we calculate
the initial distance assuming no extinction. Then the $A_{K_s}$ is
extracted from the ``mwdust'' maps for a given location ($l$, $b$, $d$)
and converted to mid-IR extinction using the mid-IR extinction curve
based on results from the ``Apache Point Observatory Galactic Evolution
Experiment'' (APOGEE) survey\cite{xue2016}
($A_{I1}/A_{K_s}=0.553$, 
$A_{I2}/A_{K_s}=0.461$, 
$A_{I3}/A_{K_s}=0.389$, 
$A_{I4}/A_{K_s}=0.426$, 
$A_{W1}/A_{K_s}=0.591$, 
$A_{W2}/A_{K_s}=0.463$, 
$A_{W3}/A_{K_s}=0.537$, 
$A_{W4}/A_{K_s}=0.364$). 
The extinction-corrected distances in all bands are averaged as
described above.  We use the new distance $d$ to extract the new value
of $A_{K_s}$ extinction from the ``mwdust'' maps. After a couple of
iterations, the process converges.

\vspace{10pt}

Table~\ref{tab:main} provides Galactic coordinates ($l$, $b$), the final
distance value and its uncertainty, estimated age, pulsation period and
mode, mean magnitudes in all eight mid-IR bands and their estimated
uncertainties (Data S1 only), as well as near-IR extinction $A_{K_s}$,
for all Cepheids in our sample.

Having the final distances obtained with the optimized mid-IR
extinction, we verified their reliability. Our distances were compared
with distances from the Gaia DR2 sample\cite{brown2018}, by selecting
counterparts with parallax errors $<\!\!10$\%. 251 common objects were
found. The comparison indicates the presence of a bias in the Gaia
parallaxes of $-0.071 \pm 0.038$ mas\cite{mroz2018}.
Ref.\cite{riess2018} noted such a bias at the level of $-0.046\pm0.013$
mas in their Cepheid sample. Other comparisons:
quasars\cite{lindegren2018}, RGB stars\cite{zinn2018}, eclipsing
binaries\cite{stassun2018}, RR Lyrae stars\cite{muraveva2018} also
suggest a bias of the Gaia DR2 distance scale at a similar level (from
$-0.030$ to $-0.082$~mas). Thus, our distance scale seems to be
compatible with current distance scale determinations. Even if our
distance scale is slightly incorrect, our picture of the Milky remains
the same -- it will just be somewhat rescaled.

The Gaia Cepheids are nearby objects. Ref.\cite{feast2014} measured
distances and extinction to five OGLE Cepheids from independent $JK$
photometry. These objects are located $\approx 15$ kpc behind the
Galactic center and are seen through high extinction. We have the mid-IR
photometry for only one of these objects, namely OGLE-BLG-CEP-005. Our
distance to this object is $23.06\pm0.77$~kpc while ref.\cite{feast2014}
gives $22.30$~kpc. The agreement to 3\% is reassuring and suggests that
our distance scale is precise at large distances as well.

These tests indicate that our distance scale is accurate to better than
5\%.

\subsubsection*{Cartesian coordinates}

The 3-D distribution of Cepheids in our sample is studied in the
Cartesian coordinate system with the origin at the Sun:
\vspace*{-7pt}
\begin{align}
        X &=  d \times \cos l \cos b \nonumber \\
        Y &=  d \times \sin l \cos b \\
        Z &=  d \times \sin b \nonumber
\end{align}
\vspace*{-30pt}

\noindent
where $l$ and $b$ are the Galactic coordinates of the star and $d$ is
its distance from the Sun. We adopt the distance between the Galactic
center and the Sun of $8.3$\,kpc\cite{pietrukowicz2015,gillessen2017}.

\subsubsection*{Galactic warp}

The 3-D distribution of Cepheids deviates from a plane. To guide the
eye, we fitted a simple polynomial surface to the distribution of
Cepheids:
\vspace*{-7pt}
\begin{align}
Z(R,\phi) &= -z_0 \qquad & {\rm for} \qquad R < R_d \nonumber \\
Z(R,\phi) &= -z_0 + z_1  (R-R_d)^2  \sin(\phi-\phi_0) \qquad  & {\rm for} \qquad  R \geq R_d
\end{align}
\vspace*{-30pt}

\noindent
where $z_0$ is the distance of the Sun from the Galactic plane, $z_1$ --
warp amplitude parameter, $R_d$ -- the distance from the Galactic center
where the warp begins and $\phi_0$ -- azimuth of the line of nodes.

We find the best-fitting parameters by minimizing the sum of squares of
orthogonal distances from the data points to the surface $\chi^2 =
\sum_i \|\vec{R}_i-\vec{R}_{\rm model}\|^2$. The best-fitting parameters
are $z_0 = 15.3$~pc (in good agreement with the Sun's height from the
thin disk exponential model -- see below), $z_1 = +0.0152$~kpc$^{-1}$
and $\phi_0=-28^{\circ}$ for fixed $R_d=8$~kpc. The best-fitting surface
is shown in Fig.~\ref{fig:warp}\,C, D, E.

We can see the Galactic warp in our 3-D map of the Galaxy using
directly measured distances to individual stars. Previous studies were
mostly based on star
counts\cite{drimmel2001,lopez2002,momany2006,reyle2009,amores2017},
distribution of red clump stars\cite{lopez2002,momany2006} and
pulsars\cite{yusifov2004}, distribution of dust\cite{marshall2006} and
gas\cite{nakanishi2003,levine2006} or stellar
kinematics\cite{smart1998,poggio2018}. Our map is a global approach --
based on observations of Cepheids in the entire Galaxy -- far from the
Galactic center, and not only in the regions relatively close to the
Sun.

Previous models do not match the shape of the Galactic warp as seen in
classical Cepheids. Models of the warp that assume a linear
shape\cite{marshall2006,reyle2009,amores2017} are likely excluded, as
our data show that the inclination of the warp is not constant
(Fig.~\ref{fig:warp}\,B). The height of the warp in the intermediate-age
population ($1.7-1.8$~kpc at $R=15$~kpc)\cite{drimmel2001,lopez2002} is
much higher than for Cepheids ($0.74$~kpc at $R=15$~kpc). The position
angle of the line of nodes of the warp is $\phi_0=-28^{\circ}$ for
Cepheids (Fig.~\ref{fig:warp}\,A), while it is closer to zero for older
stars\cite{drimmel2001,lopez2002}. This may indicate that the shape of
the warp differs between the young and older stellar populations (as
also suggested by ref.~\cite{amores2017}).

Fig.~\ref{fig:warp} also excludes the interpretation of far disk
Cepheids by ref. \cite{feast2014}. They were unaware of the large
warping of the disk toward positive latitudes in the direction of the
Galactic center ($165^{\circ}<\phi< 210^{\circ}$). Therefore, the
location of these Cepheids is typical -- in the middle of the disk --
not at the $\sim 3$-sigma edge of the flaring disk.

To illustrate the flaring of the disk, we divided it into 10 equal
sectors in the Galactocentric polar coordinate system (similar to those
in Fig.~\ref{fig:warp}), calculated the median $Z$ in 20 bins in
Galactocentric radius (provided that the bin contained at least five
stars), and subtracted the median warping.  The results are shown in
Fig.~\ref{fig:flaring}, which presents the cross-section of the entire
disk in the young population. We found that the thickness of the disk
seen in classical Cepheids is in agreement with observations of atomic
hydrogen\cite{kalberla2007}.

\subsubsection*{Disk scale height}

We model the vertical distribution of Cepheids using a simple
exponential  model of the thin disk, which has two free parameters: the
scale height  $H$ and the Sun's height above the Galactic plane $z_0$.
The density  of stars varies as:
\begin{equation}
n(z) = \frac{1}{2H} \exp\left(-\frac{|z|}{H}\right)
\end{equation}
where $z=z_0 + d\sin b$, $d$ is the distance to the Cepheid and $b$ is
its Galactic latitude. The best-fitting model is found by maximizing the
likelihood function, $\mathcal{L}$, defined as $\ln\mathcal{L}=\sum_{i}
\ln n(z_i)$. The summation is performed over all stars within 8\,kpc of
the Galactic center (to minimize the effect of the disk's warp) and
Galactic latitudes $|b|\leq 4^{\circ}$. The best-fitting parameters are
$H=73.5 \pm 3.2$\,pc and $z_0 = 14.5 \pm 3.0$\,pc. The uncertainties are
estimated using the Markov chain Monte Carlo technique\cite{foreman2013}
and represent 68\% confidence range of marginalized posterior
distributions. We assume uniform priors on $H$ and $z_0$ over the ranges
0 to 1 kpc and $-1$ to 1 kpc, respectively.

\begin{figure}[htb]
\begin{center}
\includegraphics[width=0.7\textwidth]{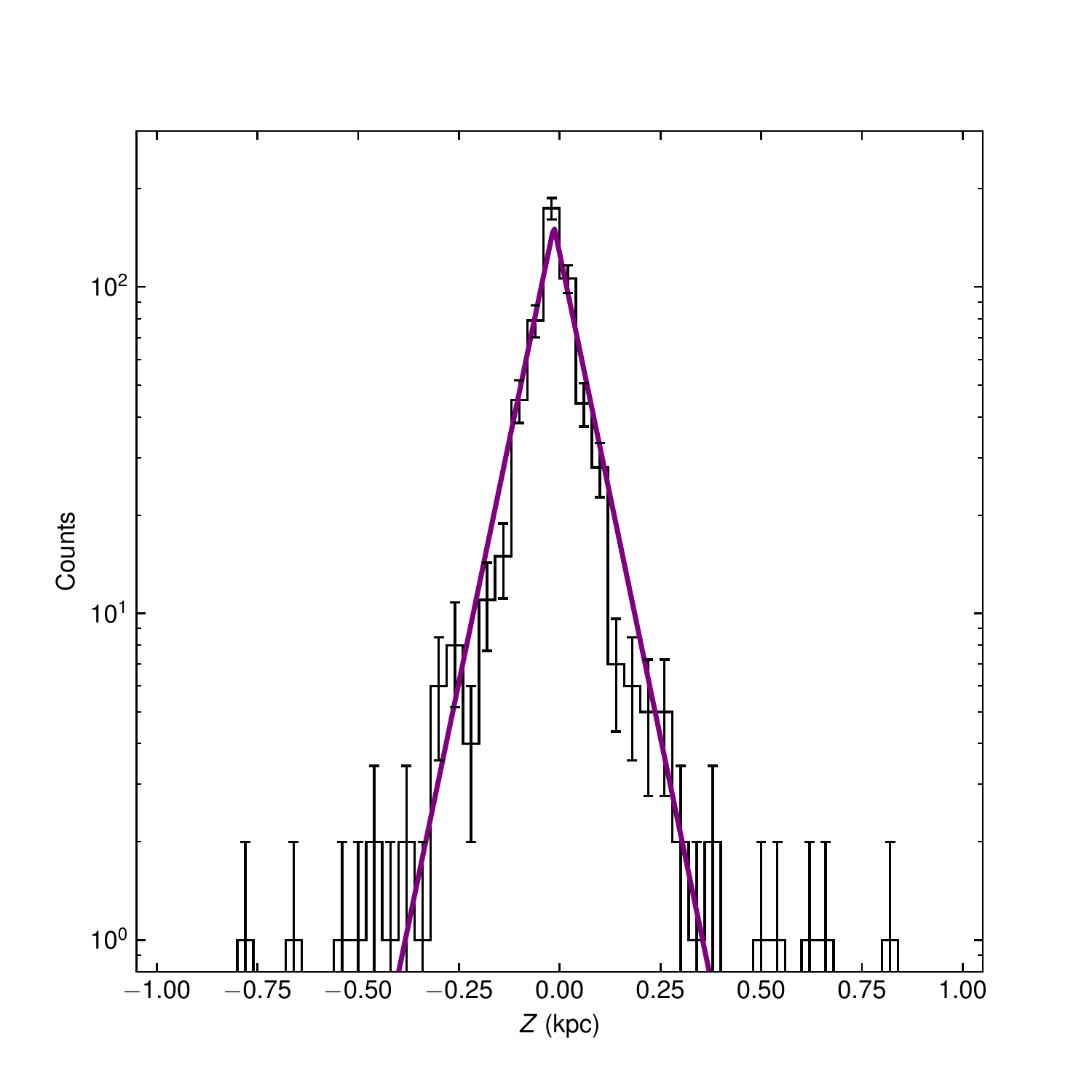}
\caption{\textbf{Histogram of distances of Cepheids from the Galactic
plane.} Histogram for objects located within 8\,kpc of the Galactic
center and  $|b|\leq 4^{\circ}$. The purple line marks the best-fitting
exponential model.}
\label{fig:hist}
\end{center}
\end{figure}

The estimated mean scale height is consistent with our measurements of 
flaring of the disk (Fig.~\ref{fig:flaring}), as the mean 
$\mathrm{HWHM}=H\ln 2 = 51$~pc. The measured parameters are in general
agreement with previous determinations (Table~\ref{tab:h_z_fit}).  The
histogram of distance $z$ together with the fit model is presented in 
Fig.~\ref{fig:hist}.

\subsubsection*{Period--Galactocentric Distance Distribution}

\begin{figure}[htb]
\begin{center}
\includegraphics[width=0.7\textwidth]{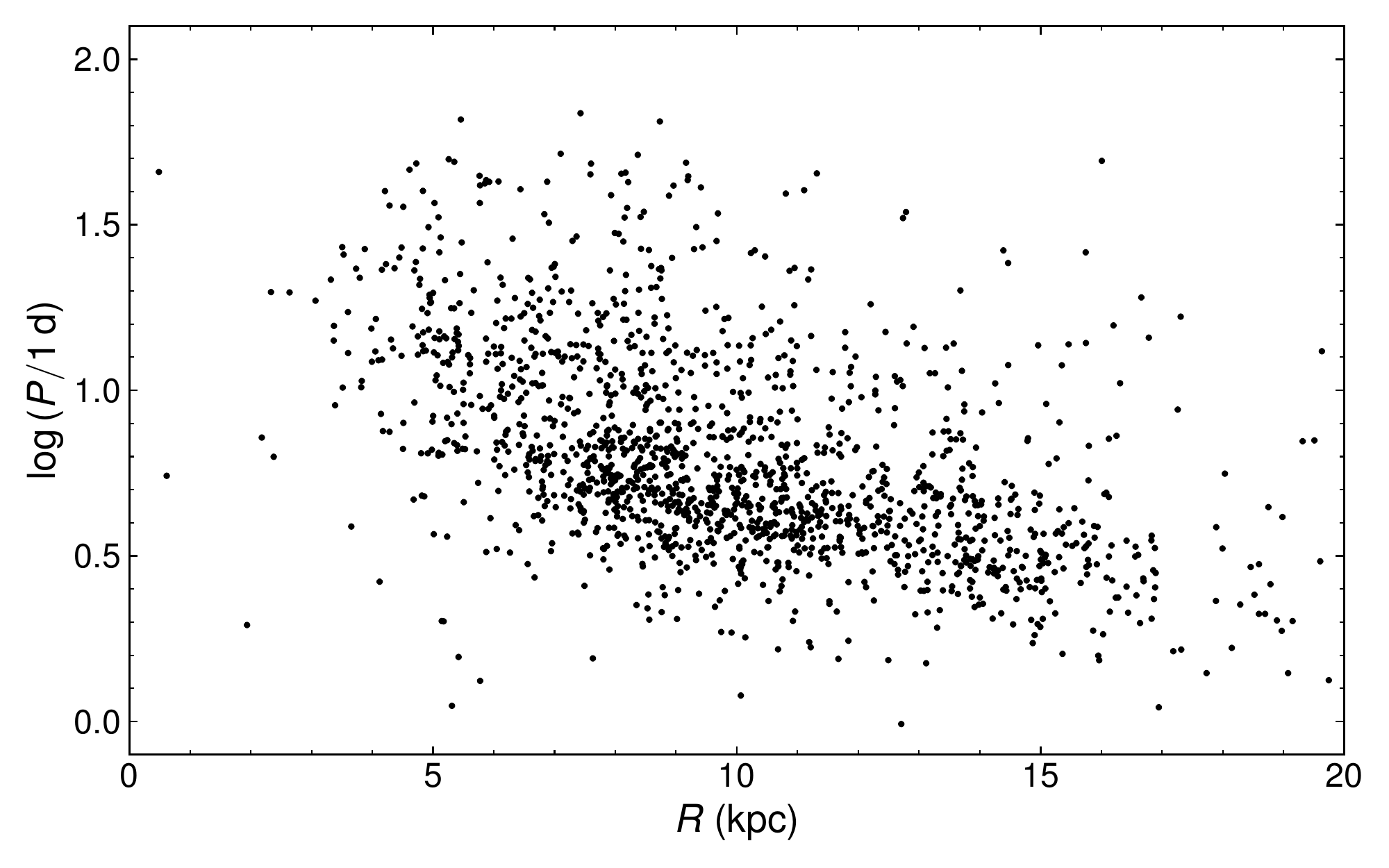}
\caption{\textbf{Distribution of periods of fundamental mode classical
Cepheids in the Milky Way as a function of the Galactocentric distance.}}
\label{fig:per-dist}
\end{center}
\end{figure}

Fig.~\ref{fig:per-dist} shows the distribution of periods of fundamental
mode Cepheids from our sample as a function of the Galactocentric
distance. A clear decrease of the minimum Cepheid period with the
distance implies a radial metallicity gradient in the Milky Way. The
metallicity gradient is also directly observed in Cepheid
spectra\cite{genovali2014}. Less massive stars in more metal rich
environments cannot reach the Cepheid instability strip in the
helium-burning phase of evolution and cannot appear then as short-period
Cepheids\cite{anderson2016}.

We carried out a simple simulation to check if a constant star formation
rate (SFR) can reproduce Fig.~\ref{fig:per-dist} at distances $R>8$~kpc,
i.e., covered by available metallicity of
models\cite{georgy2013,anderson2016}. We assumed an exponential
distribution of stellar density ($n(R)\propto\exp(-R/H_R)$) in the
Galaxy with a scale length $H_R=3$~kpc and drew stars from this
distribution. Metallicities were taken from the metallicity--distance
relation\cite{genovali2014} and their masses, $M$, and ages were
assigned from the Initial Mass Function (IMF) ($dN/dM \sim M^{-2}$) and
flat age distribution in the range of $[0,300]$ million years.  Then, we
followed the evolution of each star using evolutionary tracks
\cite{georgy2013} and if a star of a given mass, metallicity and age was
within the instability strip \cite{anderson2016}, we treated it as a
Cepheid. We assigned its present period using the period--age
relation\cite{anderson2016}.

\begin{figure}[htb] \begin{center}
\includegraphics[width=0.85\textwidth]{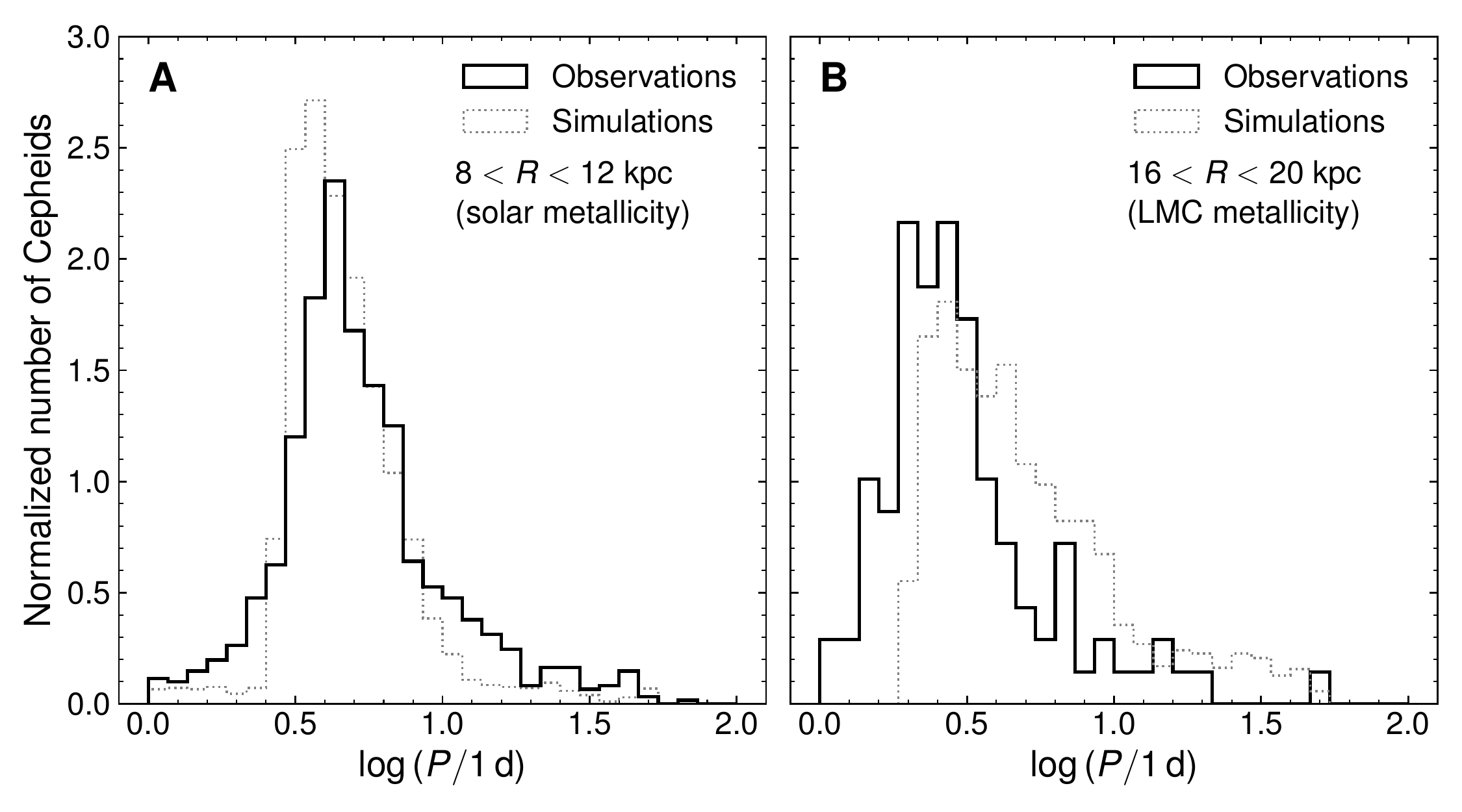}
\caption{\textbf{Observed and simulated distribution of periods of
classical Cepheids in the Milky Way, assuming a constant star formation
rate.} Panel (A) shows the distribution of periods at the solar distance
(metallicity equal to 0.014) while the panel (B) -- at the outskirts of
the Milky Way (metallicity equal to 0.006).}
\label{fig:per-hists}
\end{center}
\end{figure}

The minimum periods of our simulated Cepheid sample match the observed
values:\break $\log{P_{\rm min}}\approx 0.5$ at the distance of
$R\approx9.5$~kpc (solar metallicity of about 0.014), and $\log{P_{\rm
min}}\approx 0.3$ at the distance of $R\approx18$~kpc (metallicity of
about 0.006). However, there are fewer long-period ($\log{P} > 0.5$)
Cepheids observed in the regions far from the Galactic center
($16<R<20$~kpc) than predicted by the simulation. On the other hand, in
the regions of solar metallicity ($8<R<12$~kpc) we obtain a reverse
situation -- the number of the observed long-period Cepheids ($\log{P} >
0.8$) is higher than in the simulations (Fig.~\ref{fig:per-hists}).
These results indicate that our assumption of constant SFR is likely not
correct.  Moreover, the simulated SFR per unit volume in the outer parts
of the Galaxy would need to be lower by a factor of $4-5$ than in the
solar neighborhood to match the observations. This conclusion is in
agreement with the SFR estimated from the distribution of molecular
gas\cite{kennicutt2012}.

\subsubsection*{Age of Galactic Cepheids}

The age of a classical Cepheid is correlated with its period and the
Period-Age relation depends on metallicity\cite{bono2005}. Models show
that this relation also depends on the rotation of the
star\cite{anderson2016}. The rotation increases the age of Cepheids by a
factor of up to two. We used the Period-Age relations
\cite{anderson2016} to derive ages of Galactic Cepheids from our sample.
Because the progenitors of classical Cepheids -- B/A-type main sequence
stars -- are typically fast rotators\cite{huang2010,zorec2012} we
assumed the Period-Age relations for $\omega = 0.5$ (where $\omega$ is
the ratio of the initial angular velocity of the star to the critical
one). For a wide range of rotation values -- from moderate to very fast
($0.4 < \omega < 0.9$), i.e., applicable to the Cepheid progenitors, the
Period-Age relations are almost identical in the Galactic metallicity
range (\cite{anderson2016}, their figure~10).

Unfortunately, the available models and relations\cite{anderson2016}
were calculated for only three metallicities: solar (0.014), the Large
Magellanic Cloud  (0.006) and the Small Magellanic Cloud (0.002).
However, many Galactic Cepheids are more metal-rich than the Sun. Thus,
we first extrapolated the relations from ref.\cite{anderson2016} to
metallicity of 0.030 to cover the full range of Galactic Cepheid
metallicities. Over the whole range of Cepheid periods the more
metal-rich Cepheids are younger than the less metal-rich ones at the
same pulsation period.

There is a radial metallicity gradient in the Galaxy\cite{genovali2014}.
For each Cepheid from our sample we estimated its typical metallicity
based on the distance $R$ from the center. Then we applied the
appropriate (for fundamental-mode or first-overtone pulsations and
$\omega = 0.5$) Period-Age relation\cite{anderson2016} to derive its
age.

Results of our age determination for Galactic Cepheids are presented in
Fig.~\ref{fig:ages}\,A where the age tomography of {\rm the Milky Way
Cepheids} is plotted and Fig.~\ref{fig:ages}\,B shows the distribution
of ages of Cepheids from our sample.

\subsubsection*{Modeling the overdensities}                           

We divide the Cepheid sample into three age bins: young (20--90~Myr),
intermediate (90--140~Myr), and old (140--260~Myr) --
Fig.~\ref{fig:ages}\,C--E. The median ages of stars falling into the
marked regions are  64, 113, and 175 Myr, respectively, and standard
deviations are 16, 14, 33 Myr. The two youngest bins show a couple of
narrow overdensities with similar ages.

Because our age estimates are not accurate, we expect that stars that
formed together might show slightly different ages. However, the spatial
positions are known with much better accuracy. Thus, close spacial
proximity of stars of similar age strongly suggests their common origin
and indicates a small relative-velocity dispersion.

We model the overdensities in those three bins, assuming a single age
for all stars inside the bin. Disk stars in the Galaxy follow a flat
rotation curve with the velocity of 223 km/s\cite{mroz2018}. We take the
rotational period of the spiral pattern equal to
250~Myr\cite{vallee2017}. The age bins are illustrated in
Fig.~\ref{fig:rotation}\,A,\ D,\ G.  For each age group, we pick regions
within neighboring spiral arms, populate them randomly with stars and
assign them the rotation velocities with an additional (8, 8) km/s
dispersion in the ($U$, $V$) directions. Positions of these Cepheid
candidates are shown in Fig.~\ref{fig:rotation}\,B,\ E,\ H. Rotation of
the Galaxy to the present day is shown in Fig.~\ref{fig:rotation}\,C,\
F,\ I. We use a simple four-arm spiral structure model with a pitch
angle of 12.4 degrees\cite{vallee2017,koo2017}.

\newpage

\vspace{10pt}

\newpage

\begin{sidewaystable}

\begin{center}

\caption{\textbf{Classical Cepheid sample.} The following columns
contain: (1) Source (references only in Data~S1 table), (2) Cepheid ID,
(3,4) Galactic coordinates, (5,6) distance to the Cepheid and its error
in pc, (7) Cepheid age in Myr, (8) Pulsation period in days, (9)
Pulsation mode: fundamental (F), first-overtone (1O), (10--13) Spitzer
photometry, (14--17) WISE photometry, (18) interstellar extinction in
the $K_s$-band, (19) Source\,$''$: OGLE Galactic Cepheid Collection star
in other databases. ``$-1$'' denotes unavailable/missing data. The full
Table~S1 (2431 Cepheid entries) is available as Data~S1 table. Errors of
Spitzer and WISE magnitudes are available only in Data~S1 table.}

\vskip 8pt
\tiny
\begin{tabular}{|l@{\hspace{6pt}}l|r@{\hspace{6pt}}r@{\hspace{6pt}}r@{\hspace{6pt}}r@{\hspace{6pt}}|r@{\hspace{6pt}}r@{\hspace{6pt}}r@{\hspace{6pt}}@{\hspace{6pt}}|r@{\hspace{5pt}}r@{\hspace{5pt}}r@{\hspace{5pt}}r@{\hspace{5pt}}|r@{\hspace{5pt}}r@{\hspace{5pt}}r@{\hspace{5pt}}r@{\hspace{5pt}}|r|c|}
 
\hline
\multicolumn{1}{|c}{Source}&\multicolumn{1}{c|}{Cepheid ID}&\multicolumn{1}{c}{$l$}&\multicolumn{1}{c}{$b$}&\multicolumn{1}{c}{$d$}&\multicolumn{1}{c|}{$\sigma_d$}&\multicolumn{1}{c}{Age}&\multicolumn{1}{c}{Period}&
\multicolumn{1}{c|}{Mode}&\multicolumn{1}{c}{I1}&\multicolumn{1}{c}{I2}&\multicolumn{1}{c}{I3}&\multicolumn{1}{c|}{I4}&\multicolumn{1}{c}{W1}&\multicolumn{1}{c}{W2}&\multicolumn{1}{c}{W3}&\multicolumn{1}{c|}{W4}&
\multicolumn{1}{c|}{$A_{K_s}$}&\multicolumn{1}{c|}{Source\,$''$}\\
& &\multicolumn{1}{c}{(deg)}&\multicolumn{1}{c}{(deg)}&\multicolumn{1}{c}{(pc)}&\multicolumn{1}{c|}{(pc)}&\multicolumn{1}{c}{(Myr)}&\multicolumn{1}{c}{(d)}& &\multicolumn{1}{|c}{(mag)}&\multicolumn{1}{c}{(mag)}&
\multicolumn{1}{c}{(mag)}&\multicolumn{1}{c|}{(mag)}&\multicolumn{1}{c}{(mag)}&\multicolumn{1}{c}{(mag)}&\multicolumn{1}{c}{(mag)}&\multicolumn{1}{c|}{(mag)}&\multicolumn{1}{c|}{(mag)}&\\

\hline
OGLE-IV             & OGLE-BLG-CEP-065      & $   0.14177$ & $   1.28542$ & $ 19331$ & $  865$ & $ 130$ & $   4.5186140$ & F  & $12.202$ & $12.174$ & $12.209$ & $11.801$ & $-1.000$ & $-1.000$ & $-1.000$ & $-1.000$ & $0.670$ & -- \\
OGLE-III            & OGLE-BLG-CEP-018      & $   0.18909$ & $  -2.67713$ & $    -1$ & $    0$ & $ 101$ & $   2.8467837$ & F  & $-1.000$ & $-1.000$ & $-1.000$ & $-1.000$ & $-1.000$ & $-1.000$ & $-1.000$ & $-1.000$ & $-1.000$ & -- \\
OGLE-IV             & OGLE-BLG-CEP-078      & $   0.24164$ & $  -1.30227$ & $    -1$ & $    0$ & $ 359$ & $   0.3247108$ & 1O & $-1.000$ & $-1.000$ & $-1.000$ & $-1.000$ & $-1.000$ & $-1.000$ & $-1.000$ & $-1.000$ & $-1.000$ & -- \\
OGLE-IV             & OGLE-BLG-CEP-063      & $   0.35541$ & $   1.67398$ & $ 18326$ & $ 1254$ & $ 217$ & $   1.3795193$ & 1O & $13.266$ & $12.977$ & $-1.000$ & $-1.000$ & $-1.000$ & $-1.000$ & $-1.000$ & $-1.000$ & $0.510$ & -- \\
OGLE-IV             & OGLE-BLG-CEP-059      & $   0.38447$ & $   1.96189$ & $  6275$ & $  555$ & $ 249$ & $   0.3494682$ & 1O & $12.754$ & $13.160$ & $12.653$ & $-1.000$ & $-1.000$ & $-1.000$ & $-1.000$ & $-1.000$ & $0.445$ & -- \\
OGLE-IV             & OGLE-BLG-CEP-098      & $   0.47060$ & $  -6.65564$ & $    -1$ & $    0$ & $ 267$ & $   0.6706528$ & F  & $-1.000$ & $-1.000$ & $-1.000$ & $-1.000$ & $-1.000$ & $-1.000$ & $-1.000$ & $-1.000$ & $-1.000$ & -- \\
OGLE-IV             & OGLE-BLG-CEP-058      & $   0.50243$ & $   2.11348$ & $ 20887$ & $  696$ & $ 113$ & $   4.4882058$ & 1O & $11.656$ & $11.765$ & $11.702$ & $11.745$ & $-1.000$ & $-1.000$ & $-1.000$ & $-1.000$ & $0.513$ & -- \\
OGLE-IV             & OGLE-BLG-CEP-082      & $   0.78516$ & $  -1.25514$ & $    -1$ & $    0$ & $ 154$ & $   1.1074519$ & 1O & $-1.000$ & $-1.000$ & $-1.000$ & $-1.000$ & $-1.000$ & $-1.000$ & $-1.000$ & $-1.000$ & $-1.000$ & -- \\
OGLE-III            & OGLE-BLG-CEP-024      & $   0.92608$ & $  -3.41198$ & $ 11050$ & $  653$ & $ 247$ & $   0.3554251$ & 1O & $13.901$ & $13.982$ & $-1.000$ & $-1.000$ & $-1.000$ & $-1.000$ & $-1.000$ & $-1.000$ & $0.201$ & -- \\
OGLE-IV             & OGLE-BLG-CEP-060      & $   0.93668$ & $   2.14174$ & $ 20429$ & $  997$ & $ 134$ & $   4.6386757$ & F  & $12.203$ & $12.314$ & $11.972$ & $11.873$ & $-1.000$ & $-1.000$ & $-1.000$ & $-1.000$ & $0.549$ & -- \\
OGLE-IV             & OGLE-BLG-CEP-047      & $   1.01307$ & $   3.68211$ & $    -1$ & $    0$ & $  64$ & $   3.9657345$ & 1O & $-1.000$ & $-1.000$ & $-1.000$ & $-1.000$ & $-1.000$ & $-1.000$ & $-1.000$ & $-1.000$ & $-1.000$ & -- \\
OGLE-IV             & OGLE-BLG-CEP-080      & $   1.15305$ & $  -0.92199$ & $  5022$ & $  166$ & $ 255$ & $   0.3328426$ & 1O & $12.477$ & $12.324$ & $12.411$ & $-1.000$ & $-1.000$ & $-1.000$ & $-1.000$ & $-1.000$ & $0.464$ & -- \\
GCVS                & X\_\_\_\_\_Sgr        & $   1.16628$ & $   0.20926$ & $   350$ & $   13$ & $  90$ & $   7.0130211$ & F  & $-1.000$ & $-1.000$ & $-1.000$ & $-1.000$ & $-1.000$ & $-1.000$ & $ 2.491$ & $ 2.400$ & $0.072$ & -- \\
OGLE-III            & OGLE-BLG-CEP-019      & $   1.16962$ & $  -2.11628$ & $  3430$ & $  110$ & $ 306$ & $   0.2865322$ & 1O & $11.787$ & $11.724$ & $11.721$ & $11.572$ & $-1.000$ & $-1.000$ & $-1.000$ & $-1.000$ & $0.306$ & -- \\
OGLE-III            & OGLE-BLG-CEP-020      & $   1.20440$ & $  -2.27310$ & $  7518$ & $  326$ & $ 224$ & $   0.4406742$ & 1O & $12.866$ & $12.762$ & $12.537$ & $-1.000$ & $-1.000$ & $-1.000$ & $-1.000$ & $-1.000$ & $0.277$ & -- \\
OGLE-IV             & OGLE-BLG-CEP-066      & $   1.49232$ & $   1.73999$ & $ 17634$ & $  895$ & $ 126$ & $   2.9816002$ & 1O & $11.905$ & $12.122$ & $11.880$ & $12.105$ & $-1.000$ & $-1.000$ & $-1.000$ & $-1.000$ & $0.685$ & -- \\
OGLE-III            & OGLE-BLG-CEP-027      & $   1.56067$ & $  -4.30572$ & $  7417$ & $  314$ & $ 269$ & $   0.2965274$ & 1O & $13.305$ & $13.251$ & $-1.000$ & $-1.000$ & $-1.000$ & $-1.000$ & $-1.000$ & $-1.000$ & $0.147$ & -- \\
GCVS                & W\_\_\_\_\_Sgr        & $   1.57578$ & $  -3.97956$ & $   412$ & $   18$ & $  85$ & $   7.5950300$ & F  & $-1.000$ & $-1.000$ & $-1.000$ & $-1.000$ & $-1.000$ & $-1.000$ & $ 2.675$ & $ 2.667$ & $0.024$ & -- \\
OGLE-IV             & OGLE-BLG-CEP-044      & $   1.73682$ & $   4.38241$ & $    -1$ & $    0$ & $ 240$ & $   0.5829363$ & 1O & $-1.000$ & $-1.000$ & $-1.000$ & $-1.000$ & $-1.000$ & $-1.000$ & $-1.000$ & $-1.000$ & $-1.000$ & -- \\
ASAS                & J175258-2736.1        & $   1.97820$ & $  -0.70002$ & $  1309$ & $   32$ & $ 107$ & $   4.8224613$ & F  & $-1.000$ & $-1.000$ & $ 5.946$ & $ 5.927$ & $ 5.786$ & $ 5.818$ & $ 5.959$ & $ 5.729$ & $0.171$ & -- \\
OGLE-IV             & OGLE-BLG-CEP-076      & $   2.03608$ & $   0.10207$ & $  3184$ & $   70$ & $  66$ & $  10.3209261$ & F  & $ 7.301$ & $ 7.168$ & $ 7.130$ & $ 7.133$ & $ 7.209$ & $ 7.254$ & $ 7.322$ & $ 6.165$ & $1.056$ & -- \\
OGLE-III            & OGLE-BLG-CEP-021      & $   2.10745$ & $  -1.87113$ & $ 13081$ & $  633$ & $ 262$ & $   0.7785541$ & F  & $13.661$ & $13.604$ & $-1.000$ & $-1.000$ & $-1.000$ & $-1.000$ & $-1.000$ & $-1.000$ & $0.351$ & -- \\
OGLE-IV             & OGLE-BLG-CEP-068      & $   2.15543$ & $   1.67911$ & $    -1$ & $    0$ & $ 327$ & $   0.3716836$ & 1O & $-1.000$ & $-1.000$ & $-1.000$ & $-1.000$ & $-1.000$ & $-1.000$ & $-1.000$ & $-1.000$ & $-1.000$ & -- \\
OGLE-IV             & OGLE-BLG-CEP-074      & $   2.32318$ & $   0.80838$ & $ 17226$ & $  753$ & $ 142$ & $   3.3794859$ & F  & $12.564$ & $12.391$ & $12.164$ & $-1.000$ & $-1.000$ & $-1.000$ & $-1.000$ & $-1.000$ & $0.960$ & -- \\
OGLE-III            & OGLE-BLG-CEP-029      & $   2.43825$ & $  -4.14423$ & $    -1$ & $    0$ & $ 114$ & $   2.3760846$ & F  & $-1.000$ & $-1.000$ & $-1.000$ & $-1.000$ & $-1.000$ & $-1.000$ & $-1.000$ & $-1.000$ & $-1.000$ & -- \\
OGLE-IV             & OGLE-BLG-CEP-094      & $   2.45691$ & $  -1.72783$ & $    -1$ & $    0$ & $ 330$ & $   0.3678108$ & 1O & $-1.000$ & $-1.000$ & $-1.000$ & $-1.000$ & $-1.000$ & $-1.000$ & $-1.000$ & $-1.000$ & $-1.000$ & -- \\
OGLE-IV             & OGLE-BLG-CEP-072      & $   2.68930$ & $   1.43651$ & $  9920$ & $  442$ & $ 213$ & $   0.4918163$ & 1O & $13.392$ & $13.299$ & $-1.000$ & $-1.000$ & $-1.000$ & $-1.000$ & $-1.000$ & $-1.000$ & $0.507$ & -- \\
GCVS                & V767\_\_Sgr           & $   2.69415$ & $  -0.14902$ & $  1600$ & $   42$ & $ 245$ & $   0.6703120$ & 1O & $ 8.861$ & $ 8.757$ & $ 8.753$ & $ 8.691$ & $ 8.542$ & $ 8.612$ & $ 8.281$ & $-1.000$ & $0.187$ & -- \\
GCVS                & V773\_\_Sgr           & $   2.85322$ & $  -0.52700$ & $  1479$ & $   41$ & $  96$ & $   5.7484450$ & F  & $-1.000$ & $-1.000$ & $ 6.000$ & $ 5.996$ & $ 5.979$ & $ 5.882$ & $ 6.100$ & $ 6.019$ & $0.325$ & -- \\
OGLE-IV             & OGLE-BLG-CEP-091      & $   3.11136$ & $  -1.12969$ & $ 13209$ & $  387$ & $  47$ & $  19.4385184$ & F  & $ 9.900$ & $ 9.915$ & $ 9.717$ & $ 9.648$ & $-1.000$ & $-1.000$ & $-1.000$ & $-1.000$ & $1.950$ & -- \\
OGLE-IV             & OGLE-BLG-CEP-070      & $   3.29347$ & $   2.04078$ & $    -1$ & $    0$ & $ 145$ & $   1.2101512$ & 1O & $-1.000$ & $-1.000$ & $-1.000$ & $-1.000$ & $-1.000$ & $-1.000$ & $-1.000$ & $-1.000$ & $-1.000$ & -- \\
OGLE-IV             & OGLE-BLG-CEP-067      & $   3.58965$ & $   2.99334$ & $ 20431$ & $  660$ & $ 163$ & $   2.6107091$ & 1O & $12.362$ & $12.429$ & $12.241$ & $12.443$ & $-1.000$ & $-1.000$ & $-1.000$ & $-1.000$ & $0.343$ & -- \\
OGLE-IV             & OGLE-BLG-CEP-092      & $   3.70636$ & $  -0.78884$ & $    -1$ & $    0$ & $  56$ & $   6.8131847$ & F  & $-1.000$ & $-1.000$ & $-1.000$ & $-1.000$ & $-1.000$ & $-1.000$ & $-1.000$ & $-1.000$ & $-1.000$ & -- \\
ASAS-SN             & J180520.08-265854.0   & $   3.88312$ & $  -2.77481$ & $  7536$ & $  235$ & $ 127$ & $   1.5136369$ & 1O & $11.008$ & $11.011$ & $10.873$ & $10.789$ & $-1.000$ & $-1.000$ & $-1.000$ & $-1.000$ & $0.202$ & -- \\
GCVS                & V2744\_Oph            & $   4.12746$ & $   3.71881$ & $  1353$ & $   43$ & $  97$ & $   3.7797030$ & 1O & $-1.000$ & $-1.000$ & $-1.000$ & $-1.000$ & $ 5.856$ & $ 5.761$ & $ 5.905$ & $ 5.805$ & $0.226$ & -- \\
OGLE-IV             & OGLE-BLG-CEP-073      & $   4.26016$ & $   2.18427$ & $ 22316$ & $ 1688$ & $ 228$ & $   2.2670656$ & F  & $13.249$ & $13.456$ & $-1.000$ & $-1.000$ & $-1.000$ & $-1.000$ & $-1.000$ & $-1.000$ & $0.434$ & -- \\
OGLE-III            & OGLE-BLG-CEP-003      & $   4.34633$ & $   2.88716$ & $    -1$ & $    0$ & $ 177$ & $   1.2357274$ & F  & $-1.000$ & $-1.000$ & $-1.000$ & $-1.000$ & $-1.000$ & $-1.000$ & $-1.000$ & $-1.000$ & $-1.000$ & -- \\
OGLE-III            & OGLE-BLG-CEP-002      & $   4.56861$ & $   4.84939$ & $    -1$ & $    0$ & $ 127$ & $   2.0255716$ & F  & $-1.000$ & $-1.000$ & $-1.000$ & $-1.000$ & $-1.000$ & $-1.000$ & $-1.000$ & $-1.000$ & $-1.000$ & -- \\
OGLE-IV             & OGLE-BLG-CEP-079      & $   4.61665$ & $   1.21080$ & $ 15217$ & $  628$ & $  50$ & $  18.4782769$ & F  & $ 9.727$ & $-1.000$ & $ 9.540$ & $-1.000$ & $-1.000$ & $-1.000$ & $-1.000$ & $-1.000$ & $0.756$ & -- \\
ASAS                & J162326-0941.0        & $   4.84083$ & $  26.82556$ & $  1734$ & $   51$ & $ 166$ & $   1.3782385$ & 1O & $-1.000$ & $-1.000$ & $-1.000$ & $-1.000$ & $ 7.796$ & $ 7.819$ & $ 7.772$ & $ 0.000$ & $0.082$ & -- \\
\ldots              & \ldots                & \ldots       & \ldots       & \ldots   & \ldots  & \ldots & \ldots        &\ldots&\ldots   & \ldots   & \ldots   & \ldots   & \ldots   & \ldots   & \ldots   & \ldots   & \ldots & \ldots \\
GCVS                & V1828\_Sgr             & $   5.28401$ & $  -4.10021$ & $ 16825$ & $  941$ & $  64$ & $  12.9751466$ & F  & $-1.000$ & $-1.000$ & $-1.000$ & $-1.000$ & $10.010$ & $10.171$ & $10.302$ & $-1.000$ & $-1.000$ & OGLE \\
\ldots              & \ldots                & \ldots       & \ldots       & \ldots   & \ldots  & \ldots & \ldots        &\ldots&\ldots   & \ldots   & \ldots   & \ldots   & \ldots   & \ldots   & \ldots   & \ldots   & \ldots & \ldots \\
OGLE-IV             & OGLE-BLG-CEP-071      & $   6.28869$ & $   3.85637$ & $  8954$ & $  892$ & $ 144$ & $   1.1533927$ & 1O & $-1.000$ & $-1.000$ & $-1.000$ & $-1.000$ & $11.612$ & $11.939$ & $-1.000$ & $-1.000$ & $0.340$ &  GAIA \\
\ldots              & \ldots                & \ldots       & \ldots       & \ldots   & \ldots  & \ldots & \ldots        &\ldots&\ldots   & \ldots   & \ldots   & \ldots   & \ldots   & \ldots   & \ldots   & \ldots   & \ldots & \ldots \\
OGLE-IV             & OGLE-BLG-CEP-149      & $  13.23464$ & $  -0.20047$ & $  2800$ & $   97$ & $  60$ & $  12.5168351$ & F  & $-1.000$ & $-1.000$ & $ 6.162$ & $ 6.198$ & $ 6.147$ & $ 6.114$ & $ 6.381$ & $-1.000$ & $0.163$ &  ASAS-SN \\
\ldots              & \ldots                & \ldots       & \ldots       & \ldots   & \ldots  & \ldots & \ldots        &\ldots&\ldots   & \ldots   & \ldots   & \ldots   & \ldots   & \ldots   & \ldots   & \ldots   & \ldots & \ldots \\
OGLE-IV             & OGLE-BLG-CEP-150      & $  14.35262$ & $   0.34341$ & $  3463$ & $   78$ & $  64$ & $   7.1806422$ & 1O & $ 7.244$ & $ 7.072$ & $ 7.040$ & $ 6.973$ & $ 7.127$ & $ 7.100$ & $ 7.359$ & $-1.000$ & $0.504$ &  ATLAS \\
\ldots              & \ldots                & \ldots       & \ldots       & \ldots   & \ldots  & \ldots & \ldots        &\ldots&\ldots   & \ldots   & \ldots   & \ldots   & \ldots   & \ldots   & \ldots   & \ldots   & \ldots & \ldots \\
\hline
\end{tabular}
\label{tab:main}

\end{center}

\end{sidewaystable}

\begin{table}

\begin{center}

\caption{\textbf{Comparison of recent determinations of the scale height of the
thin disk ${\pmb H}$ and the Sun's height above the Galactic plane ${\pmb
z}_{\pmb 0}$.}}

\vskip 8pt

\begin{tabular}{clrr}
\hline \hline
\noalign{\vskip 3pt}
Reference & Tracers & \multicolumn{1}{c}{$H$ (pc)} & \multicolumn{1}{c}{$z_0$ (pc)} \\
\hline
\noalign{\vskip 3pt}
This study                  	& Cepheids                                   & $73.5 \pm 3.2$  & $14.5 \pm 3.0$ \\
\hline
\noalign{\vskip 3pt}
\cite{karim2017}		& Sgr A* offset from the Galactic plane      & & $17.1 \pm 5.0$ \\
\cite{yao2017}		  	& Pulsars                                    & $56.9 \pm 6.5$ & $13.4 \pm 4.4$  \\
\cite{joshi2016}	        & Open clusters                              & $64 \pm 2$ & $6.2 \pm 1.1$ \\
\cite{bobylev2016b}		& Masers                                     & $24.1 \pm 0.9$ & $5.7 \pm 0.5$ \\
\cite{bobylev2016a}		& Cepheids                                   & $66.2 \pm 1.6$ & $16 \pm 2$ \\
\cite{olausen2014}		& Magnetars                                  & $30.7 \pm 5.9$ & $13.5 \pm 2.6$ \\
\cite{buckner2014}		& Open clusters                              & $\sim 40$ & $18.5 \pm 1.2$ \\
\cite{liu2011}			& Open clusters                              & $58 \pm 4$ & $16 \pm 4$ \\
\cite{majaess2009}		& Cepheids                                   & $<70 \pm 10$ & $26 \pm 3$ \\
\cite{juric2008}		& photometric parallaxes		     & $300 \pm 60$ & $25 \pm 5$ \\
\cite{chen2001}			& SDSS star counts                           & $330 \pm 3$ & $27 \pm 4$ \\
\hline
\end{tabular}
\label{tab:h_z_fit}
\end{center}
\end{table}


\begin{thebibliography}{10}

\bibitem{leavitt1912}
H.~S. {Leavitt}, E.~C. {Pickering}, {\it Harvard College Observatory
  Circular\/} {\bf 173}, 1 (1912).

\bibitem{udalski2015}
A.~{Udalski}, M.~K. {Szyma{\'n}ski}, G.~{Szyma{\'n}ski}, {\it \actaa\/} {\bf
  65}, 1 (2015).

\bibitem{udalski2018}
A.~{Udalski}, {\it et~al.\/}, {\it \actaa\/} {\bf 68}, 315 (2018).

\bibitem{pietrukowicz2013}
P.~{Pietrukowicz}, {\it et~al.\/}, {\it \actaa\/} {\bf 63}, 379 (2013).

\bibitem{samus2017}
N.~N. {Samus}, E.~V. {Kazarovets}, O.~V. {Durlevich}, N.~N. {Kireeva}, E.~N.
  {Pastukhova}, {\it Astronomy Reports\/} {\bf 61}, 80 (2017).

\bibitem{pojmanski2002}
G.~{Pojma{\'n}ski}, {\it \actaa\/} {\bf 52}, 397 (2002).

\bibitem{asassn}
T.~{Jayasinghe}, {\it et~al.\/}, {\it \mnras\/} {\bf 477}, 3145 (2018).

\bibitem{atlas}
A.~N. {Heinze}, {\it et~al.\/}, {\it \aj\/} {\bf 156}, 241 (2018).

\bibitem{holl2018}
B.~{Holl}, {\it et~al.\/}, {\it \aap\/} {\bf 618}, A30 (2018).

\bibitem{suppl}
Materials and methods are available as supplementary materials.

\bibitem{wang2018}
S.~{Wang}, X.~{Chen}, R.~{de Grijs}, L.~{Deng}, {\it \apj\/} {\bf 852}, 78
  (2018).

\bibitem{levine2006}
E.~S. {Levine}, L.~{Blitz}, C.~{Heiles}, {\it Science\/} {\bf 312}, 1773
  (2006).

\bibitem{nakanishi2003}
H.~{Nakanishi}, Y.~{Sofue}, {\it \pasj\/} {\bf 55}, 191 (2003).

\bibitem{drimmel2001}
R.~{Drimmel}, D.~N. {Spergel}, {\it \apj\/} {\bf 556}, 181 (2001).

\bibitem{lopez2002}
M.~{L{\'o}pez-Corredoira}, A.~{Cabrera-Lavers}, F.~{Garz{\'o}n}, P.~L.
  {Hammersley}, {\it \aap\/} {\bf 394}, 883 (2002).

\bibitem{momany2006}
Y.~{Momany}, {\it et~al.\/}, {\it \aap\/} {\bf 451}, 515 (2006).

\bibitem{reyle2009}
C.~{Reyl{\'e}}, D.~J. {Marshall}, A.~C. {Robin}, M.~{Schultheis}, {\it \aap\/}
  {\bf 495}, 819 (2009).

\bibitem{amores2017}
E.~B. {Am{\^o}res}, A.~C. {Robin}, C.~{Reyl{\'e}}, {\it \aap\/} {\bf 602}, A67
  (2017).

\bibitem{marshall2006}
D.~J. {Marshall}, A.~C. {Robin}, C.~{Reyl{\'e}}, M.~{Schultheis}, S.~{Picaud},
  {\it \aap\/} {\bf 453}, 635 (2006).

\bibitem{smart1998}
R.~L. {Smart}, R.~{Drimmel}, M.~G. {Lattanzi}, J.~J. {Binney}, {\it \nat\/}
  {\bf 392}, 471 (1998).

\bibitem{poggio2018}
E.~{Poggio}, {\it et~al.\/}, {\it \mnras\/} {\bf 481}, L21 (2018).

\bibitem{kalberla2007}
P.~M.~W. {Kalberla}, L.~{Dedes}, J.~{Kerp}, U.~{Haud}, {\it \aap\/} {\bf 469},
  511 (2007).

\bibitem{antonello2002}
E.~{Antonello}, D.~{Fugazza}, L.~{Mantegazza}, M.~{Stefanon}, S.~{Covino}, {\it
  \aap\/} {\bf 386}, 860 (2002).

\bibitem{anderson2016}
R.~I. {Anderson}, H.~{Saio}, S.~{Ekstr{\"o}m}, C.~{Georgy}, G.~{Meynet}, {\it
  \aap\/} {\bf 591}, A8 (2016).

\bibitem{genovali2014}
K.~{Genovali}, {\it et~al.\/}, {\it \aap\/} {\bf 566}, A37 (2014).

\bibitem{majaess2009}
D.~J. {Majaess}, D.~G. {Turner}, D.~J. {Lane}, {\it \mnras\/} {\bf 398}, 263
  (2009).

\bibitem{dambis2015}
A.~K. {Dambis}, {\it et~al.\/}, {\it Astronomy Letters\/} {\bf 41}, 489 (2015).

\bibitem{sanna2017}
A.~{Sanna}, M.~J. {Reid}, T.~M. {Dame}, K.~M. {Menten}, A.~{Brunthaler}, {\it
  Science\/} {\bf 358}, 227 (2017).

\bibitem{vallee2017}
J.~P. {Vall{\'e}e}, {\it The Astronomical Review\/} {\bf 13}, 113 (2017).

\bibitem{mroz2018}
P.~{Mr{\'o}z}, {\it et~al.\/}, {\it \apjl\/} {\bf 870}, L10 (2019).

\bibitem{soszynski2017}
I.~{Soszy{\'n}ski}, {\it et~al.\/}, {\it \actaa\/} {\bf 67}, 297 (2017).

\bibitem{soszynski2011}
I.~{Soszy{\'n}ski}, {\it et~al.\/}, {\it \actaa\/} {\bf 61}, 285 (2011).

\bibitem{soszynski2008}
I.~{Soszy{\'n}ski}, {\it et~al.\/}, {\it \actaa\/} {\bf 58}, 163 (2008).

\bibitem{wise}
E.~L. {Wright}, {\it et~al.\/}, {\it \aj\/} {\bf 140}, 1868 (2010).

\bibitem{neowise}
A.~{Mainzer}, {\it et~al.\/}, {\it \apj\/} {\bf 731}, 53 (2011).

\bibitem{klagyivik2009}
P.~{Klagyivik}, L.~{Szabados}, {\it \aap\/} {\bf 504}, 959 (2009).

\bibitem{glimpseI}
R.~A. {Benjamin}, {\it et~al.\/}, {\it \pasp\/} {\bf 115}, 953 (2003).

\bibitem{glimpse}
E.~{Churchwell}, {\it et~al.\/}, {\it \pasp\/} {\bf 121}, 213 (2009).

\bibitem{bovy2016}
J.~{Bovy}, H.-W. {Rix}, G.~M. {Green}, E.~F. {Schlafly}, D.~P. {Finkbeiner},
  {\it \apj\/} {\bf 818}, 130 (2016).

\bibitem{green2015}
G.~M. {Green}, {\it et~al.\/}, {\it \apj\/} {\bf 810}, 25 (2015).

\bibitem{saito2010}
R.~K. {Saito}, {\it et~al.\/}, {\it \aap\/} {\bf 537}, A107 (2012).

\bibitem{xue2016}
M.~{Xue}, {\it et~al.\/}, {\it \apjs\/} {\bf 224}, 23 (2016).

\bibitem{brown2018}
{Gaia Collaboration}, {\it et~al.\/}, {\it \aap\/} {\bf 616}, A1 (2018).

\bibitem{riess2018}
A.~G. {Riess}, {\it et~al.\/}, {\it \apj\/} {\bf 861}, 126 (2018).

\bibitem{lindegren2018}
L.~{Lindegren}, {\it et~al.\/}, {\it \aap\/} {\bf 616}, A2 (2018).

\bibitem{zinn2018}
J.~C. {Zinn}, M.~H. {Pinsonneault}, D.~{Huber}, D.~{Stello}, {\it
\apj\/} {\bf 878}, 136 (2019).

\bibitem{stassun2018}
K.~G. {Stassun}, G.~{Torres}, {\it \apj\/} {\bf 862}, 61 (2018).

\bibitem{muraveva2018}
T.~{Muraveva}, H.~E. {Delgado}, G.~{Clementini}, L.~M. {Sarro}, A.~{Garofalo},
  {\it \mnras\/} {\bf 481}, 1195 (2018).

\bibitem{feast2014}
M.~W. {Feast}, J.~W. {Menzies}, N.~{Matsunaga}, P.~A. {Whitelock}, {\it \nat\/}
  {\bf 509}, 342 (2014).

\bibitem{pietrukowicz2015}
P.~{Pietrukowicz}, {\it et~al.\/}, {\it \apj\/} {\bf 811}, 113 (2015).

\bibitem{gillessen2017}
S.~{Gillessen}, {\it et~al.\/}, {\it \apj\/} {\bf 837}, 30 (2017).

\bibitem{yusifov2004}
I.~{Yusifov}, {\it The Magnetized Interstellar Medium\/}, B.~{Uyaniker},
  W.~{Reich}, R.~{Wielebinski}, eds. (2004), pp. 165--169.

\bibitem{foreman2013}
D.~{Foreman-Mackey}, D.~W. {Hogg}, D.~{Lang}, J.~{Goodman}, {\it \pasp\/} {\bf
  125}, 306 (2013).

\bibitem{georgy2013}
C.~{Georgy}, {\it et~al.\/}, {\it \aap\/} {\bf 553}, A24 (2013).

\bibitem{kennicutt2012}
R.~C. {Kennicutt}, N.~J. {Evans}, {\it \araa\/} {\bf 50}, 531 (2012).

\bibitem{bono2005}
G.~{Bono}, {\it et~al.\/}, {\it \apj\/} {\bf 621}, 966 (2005).

\bibitem{huang2010}
W.~{Huang}, D.~R. {Gies}, M.~V. {McSwain}, {\it \apj\/} {\bf 722}, 605 (2010).

\bibitem{zorec2012}
J.~{Zorec}, F.~{Royer}, {\it \aap\/} {\bf 537}, A120 (2012).

\bibitem{koo2017}
B.-C. {Koo}, {\it et~al.\/}, {\it \pasp\/} {\bf 129}, 094102 (2017).

\bibitem{karim2017}
M.~T. {Karim}, E.~E. {Mamajek}, {\it \mnras\/} {\bf 465}, 472 (2017).

\bibitem{yao2017}
J.~M. {Yao}, R.~N. {Manchester}, N.~{Wang}, {\it \mnras\/} {\bf 468}, 3289
  (2017).

\bibitem{joshi2016}
Y.~C. {Joshi}, A.~K. {Dambis}, A.~K. {Pandey}, S.~{Joshi}, {\it \aap\/} {\bf
  593}, A116 (2016).

\bibitem{bobylev2016b}
V.~V. {Bobylev}, A.~T. {Bajkova}, {\it Astronomy Letters\/} {\bf 42}, 182
  (2016).

\bibitem{bobylev2016a}
V.~V. {Bobylev}, A.~T. {Bajkova}, {\it Astronomy Letters\/} {\bf 42}, 1 (2016).

\bibitem{olausen2014}
S.~A. {Olausen}, V.~M. {Kaspi}, {\it \apjs\/} {\bf 212}, 6 (2014).

\bibitem{buckner2014}
A.~S.~M. {Buckner}, D.~{Froebrich}, {\it \mnras\/} {\bf 444}, 290 (2014).

\bibitem{liu2011}
J.~{Liu}, Z.~{Zhu}, {\it 9th Pacific Rim Conference on Stellar Astrophysics\/},
  S.~{Qain}, K.~{Leung}, L.~{Zhu}, S.~{Kwok}, eds. (2011), vol. 451 of {\it
  Astronomical Society of the Pacific Conference Series\/}, p. 339.

\bibitem{juric2008}
M.~{Juri{\'c}}, {\it et~al.\/}, {\it \apj\/} {\bf 673}, 864 (2008).

\bibitem{chen2001}
B.~{Chen}, {\it et~al.\/}, {\it \apj\/} {\bf 553}, 184 (2001).

\end{thebibliography}
\end{document}